\documentclass[aps,prd,twocolumn,superscriptaddress,nofootinbib]{revtex4-2}

\usepackage{amsmath,amssymb}
\usepackage{graphicx}
\usepackage{subcaption}
\usepackage{hyperref}
\usepackage{physics, yfonts, comment}
\usepackage{booktabs}
\usepackage{siunitx}
\usepackage{orcidlink}

\begin{document}

\title{High-Order Pole-Skipping in Near-Extremal Holography}

\author{Xiang Li \orcidlink{0009-0000-2039-5212}}
\affiliation{Department of Physics, College of Sciences, Shanghai University, 99 Shangda Road, Shanghai 200444, China}
\author{Haiming Yuan \orcidlink{0009-0002-0610-7936}}
\affiliation{School of Physics and Advanced Energy, Henan University of Technology, 100 Lianhua Street, Zhengzhou 450001, China}
\author{Xian-Hui Ge \orcidlink{0000-0001-6228-0376}}
\email{ gexh@shu.edu.cn (corresponding author)}
\affiliation{Department of Physics, College of Sciences, Shanghai University, 99 Shangda Road, Shanghai 200444, China}

\begin{abstract}
We develop a systematic analytic method for studying high-order pole-skipping in near-extremal holographic black holes. In the near-extremal regime, approaching the limit $T\to0$, the near-horizon geometry develops an approximately $\mathrm{AdS}_2 \times \mathbb{R}^{d-1}$ structure; we show that the mode index $q$ labeling pole-skipping points is identified with the IR conformal dimension $\Delta_{\mathrm{IR}} = q$ in the emergent $\mathrm{AdS}_2/\mathrm{CFT}_1$ correspondence, providing a concrete physical interpretation of the subleading pole-skipping tower. The method reorganizes the near-horizon Frobenius expansion according to powers of temperature. This reveals a temperature-graded hierarchical structure that reduces the $n$-th-order pole-skipping condition to a factorized algebraic equation:
each pole-skipping momentum depends only on the mode index $q$, not on the
order $n$. This $n$-independence produces a high degeneracy as $T\to 0$, where pole-skipping momenta at all orders collapse onto a discrete set of values determined by near-horizon geometry and the scalar field mass; these values can be expressed in terms of thermodynamic quantities such as entropy density and specific heat.
In the limit $n \gg 1$ (with $nT$ remaining small), the leading pole-skipping momenta grow asymptotically as $k_{n,n} \propto n$.
We compute leading temperature corrections and verify our predictions through numerical analysis of the Dyonic Gubser--Rocha model. The results confirm that high-order pole-skipping at low temperature is governed by near-horizon physics. This provides analytic access to pole-skipping points well beyond those accessible by standard determinant methods and clarifies the structure of holographic Green's functions in the low-temperature regime.
\end{abstract}

\maketitle

\section{Introduction}

The holographic correspondence between gravitational theories in asymptotically anti-de Sitter (AdS) spacetimes and strongly coupled quantum field theories (QFTs) provides a framework for computing real-time correlation functions~\cite{Maldacena:1997re,Gubser:1998bc,Witten:1998qj,Duff:1994an}.
Within this framework, retarded Green's functions of the boundary theory are encoded in the behavior of bulk perturbations subject to ingoing boundary conditions at black hole horizons.
Such holographic Green's functions exhibit a property known as pole-skipping~\cite{Grozdanov:2017ajz,Blake:2016wvh,Blake:2018leo,Blake:2017ris,Natsuume:2019sfp,Ahn:2020bks,Yuan:2020fvv}.

Pole-skipping refers to the fact that the retarded Green's function
\begin{equation}
G^R(\omega,k)=\frac{b(\omega,k)}{a(\omega,k)}
\end{equation}
becomes ill-defined at special points $(\omega_\star,k_\star)$ in the complex frequency--momentum plane where both numerator and denominator vanish simultaneously,
$a(\omega_\star,k_\star)=b(\omega_\star,k_\star)=0$.
Holographically, these special points correspond to the appearance of trivial ingoing solutions of the bulk equations of motion near the black hole horizon.

In holographic chaos, the central quantities of interest are the out-of-time-ordered correlator (OTOC) and the associated Lyapunov exponent $\lambda_L$ and butterfly velocity $v_B$. These quantities can be extracted from the retarded Green's function via pole-skipping~\cite{Grozdanov:2017ajz,Blake:2016wvh,Blake:2017ris,Choi:2020tdj,Chua:2025vig}, which provided the original motivation for studying this phenomenon. In this context, one focuses on the gravitational sound channel, or equivalently the energy-density Green's function,. While scalar and other bulk fields generically exhibit pole-skipping points in the lower-half $\omega$-plane, the gravitational sound mode possesses a distinguished pole-skipping point in the upper-half $\omega$-plane at $\omega_{\mathrm{chaos}} = +i\lambda_L = +i(2\pi T)$. At this special point, the frequency and momentum are directly related to the Lyapunov exponent and butterfly velocity, $\omega = i\lambda_L$ and $k = i\lambda_L/v_B$, a relation that can be further exploited to extract $\lambda_L$ and $v_B$ in the near-extremal limit. Recent studies have explored these connections in various settings, including critical points~\cite{Amrahi:2023xso}, $T\bar{T}$-deformed theories~\cite{Basu:2025exh}, and stringy horizons~\cite{Gao:2025ohk}. Beyond the leading chaotic point, there exists an infinite tower of subleading pole-skipping points whose physical interpretation remains an active area of investigation~\cite{Ceplak:2019ymw,Jansen:2020hfd,Yuan:2024utc,Yuan:2025ivz,Yuan:2021ets,Yuan:2020fvv}.
In this work we resolve this question in the near-extremal setting.

The near-horizon analysis of bulk perturbations provides a method to determine pole-skipping points~\cite{Blake:2017ris,Liu:2018kfw}.
By expanding the equations of motion around the black hole horizon, one obtains a set of recursion relations for the near-horizon coefficients.
Pole-skipping occurs when the corresponding recursion matrix becomes singular.
While this method is systematic in principle, its practical implementation becomes increasingly complicated at higher orders.
The $n$-th pole-skipping condition involves the vanishing of an $n\times n$ determinant constructed from near-horizon expansion coefficients.
As the order increases, the algebraic complexity grows rapidly, making analytic control difficult and often restricting the analysis to low orders or numerical treatments~\cite{Ahn:2020bks,Yuan:2023tft}.
Various approaches have been developed to classify and understand the structure of pole-skipping points~\cite{Ahn:2020baf,Lu:2025pal,Lu:2025jgk,Natsuume:2021fhn}, and extensions to rotating black holes~\cite{Blake:2021hjj,Jeong:2023rck}, two-dimensional CFTs~\cite{Ramirez:2020qer}, and massive fields~\cite{Pan:2024azf} have been explored.

The near-extremal regime of black holes provides a qualitatively different arena.
As the Hawking temperature $T$ approaches zero, the near-horizon geometry develops an enhanced structure and the coefficients in the recursion relations acquire nontrivial temperature dependence.
High-order pole-skipping points may reorganize into a simpler pattern governed by near-horizon data rather than the full bulk geometry.
However, a systematic understanding of high-order pole-skipping in the near-extremal limit has remained elusive~\cite{Natsuume:2019sfp,Ceplak:2019ymw,Abbasi:2020ykq,Blake:2019otz}.

The enhanced near-horizon structure that emerges as $T \to 0$ is not an arbitrary simplification but reflects a fundamental feature of Einstein gravity: the development of an approximately AdS$_2$ near-horizon throat. This geometry is universal, appearing in any extremal black hole solution regardless of the asymptotic boundary conditions or matter content, and is the holographic dual of a zero-temperature quantum critical point. The approximately AdS$_2$ factor carries its own infrared conformal symmetry, organizing low-energy excitations into a tower of operators; the pole-skipping condition selects those with integer IR scaling dimensions. If high-order pole-skipping is indeed controlled by near-horizon physics, the pole-skipping spectrum should reflect this emergent IR conformal structure.

In this work, we develop a systematic near-extremal expansion that naturally reveals this simplification. The essential observation is that at the Matsubara frequencies $\omega_n = -i2\pi Tn$, the superdiagonal elements of the near-horizon recursion matrix vanish linearly with temperature, $M_{j,j+1} = 2\pi T(j-n) \sim \mathcal{O}(T)$, while the diagonal elements $M_{jj}^{(0)}$ retain $\mathcal{O}(T^0)$ contributions from the near-horizon geometry. In the limit $T \to 0$, the matrix therefore becomes upper bidiagonal, and the $n \times n$ determinant factorizes as
\begin{equation}
    \det\mathcal{M}^{(n)} = \prod_{\sigma=1}^{n} M_{\sigma\sigma}^{(0)} + \mathcal{O}(T).
\end{equation}
Since each factor in the product depends on the mode index $q$ but is independent of the order $n$, the pole-skipping momenta at all orders $n \geq q$ with the same mode index $q$ are governed by the same condition and collapse to a common value as $T\to 0$,
\begin{equation}
k_{n,q}^2 \;\xrightarrow{\;T\to 0\;}\; k_q^2
= -m^2 h(r_h) + \frac{q(q-1)}{2}\,h(r_h)\,f''(r_h)\,.
\end{equation}
The $n$-independence of $k_q^2$ is the hallmark of the near-extremal regime: all pole-skipping points in a given branch $q$ are controlled by the same near-horizon geometry, regardless of their order in the Matsubara tower.

The pole-skipping momenta $k_q^2$ have a direct geometric interpretation. Substituting into the effective near-horizon AdS$_2$ mass relation in Eq.~\eqref{eq:IR-dim-early}, one finds $m_{\mathrm{eff}}^2 L_2^2 = q(q-1)$, so that the scalar field in the emergent approximately AdS$_2$ throat acquires an IR conformal dimension $\Delta_{\mathrm{IR}} = q$. The mode index $q$ thus labels the effective-mass sector of the AdS$_2$ problem set by the parent-theory momentum $k$: at the pole-skipping momentum $k_q$, the scalar matches an integer IR scaling dimension in the emergent throat. This provides a concrete physical interpretation of the subleading pole-skipping tower in terms of the AdS$_2$/CFT$_1$ conformal structure.

We further compute the leading temperature correction, obtaining $k_{n,q}^2 = k_q^2 + C_{n,q}\,T + \mathcal{O}(T^2)$, where the coefficient $C_{n,q}$ depends on both $n$ and $q$ and lifts the extremal degeneracy. The $n$-dependence of $C_{n,q}$ encodes how different orders of pole-skipping respond differently to thermal perturbations.

We verify these predictions numerically in the Dyonic Gubser--Rocha model~\cite{Ge:2023yom,Xu:2023qlu,Ishigaki:2024djz}, which incorporates finite charge density and momentum relaxation and admits a smooth extremal limit. Numerical evaluation of the full $n\times n$ determinant condition confirms the analytic predictions with relative errors at the level of $10^{-29}$--$10^{-15}$ across multiple orders $n$, mode indices $q$, and temperature regimes.

The remainder of this paper is organized as follows. Sec.~\ref{sec:near-extremal} develops the analytic framework and derives the main results. Sec.~\ref{sec:GR} presents the numerical verification in the Dyonic Gubser--Rocha model. Sec.~\ref{sec:conclusion} concludes with a discussion of open directions.

To orient the reader, the hierarchy of results is as follows. The primary result is the $n$-independent collapse formula~\eqref{eq:extremal-momenta}, $k_{n,q}^2 \to k_q^2$, which follows from the factorization of the near-horizon recursion determinant in the limit $T\to 0$, where the extremal factorization emerges. Its geometric interpretation that the mode index $q$ equals the IR conformal dimension $\Delta_{\mathrm{IR}} = q$, and the leading temperature correction Eq.~\eqref{eq:T-correction} are secondary results that follow from it.

\section{Pole-Skipping in Near-Extremal Black Holes}\label{sec:near-extremal}

The near-extremal regime $T \to 0$ is distinguished not merely by a parametric simplification but by the emergence of qualitatively new geometric structure. As we review in Sec.~\ref{sec:motivation}, the near-extremal black holes studied here are solutions to the Einstein--Maxwell--Scalar action that admit a smooth extremal limit; as $T\to 0$, the near-horizon region develops an approximately AdS$_2$ throat whose properties are to leading order determined by the near-horizon dynamics of that action.

Our primary result is that high-order pole-skipping in the near-extremal regime is governed by this emergent near-horizon structure rather than the full bulk geometry. The pole-skipping spectrum reflects the IR conformal tower of the AdS$_2$/CFT$_1$ correspondence, with the mode index $q$ corresponding to the IR scaling dimension $\Delta_{\mathrm{IR}} = q$. This section develops the systematic near-extremal expansion that makes this structure manifest: we first establish the physical motivation and the emergent AdS$_2$ geometry in Sec.~\ref{sec:motivation}, then review the standard near-horizon analysis in Sec.~\ref{sec:Standard Near-Horizon Analysis}, develop the systematic expansion method exploiting the temperature hierarchy to derive the pole-skipping momenta in the extremal limit in Secs.~\ref{sec:Systematic Near-Extremal Expansion Method}--\ref{sec:Pole-Skipping Momenta in the Near-Extremal Limit}, compute the leading temperature corrections away from extremality in Sec.~\ref{sec:leading-temp-corrections}, and verify the analytic results against known exact solutions in simple holographic backgrounds in Sec.~\ref{sec:comparison}.

\subsection{Near-Extremal Black Holes and Emergent AdS\texorpdfstring{$_2$}{2} Geometry}\label{sec:motivation}

The near-extremal black holes studied in this work arise as solutions to the Einstein--Maxwell--Scalar (EMS) action,
\begin{widetext}
\begin{equation}\label{eq:EMS-action}
    S = \frac{1}{16\pi G_N}\int d^{d+1}x\sqrt{-g}\left(R - \frac{1}{2}(\partial\phi)^2 - V(\phi) - \frac{Z(\phi)}{4}F_{\mu\nu}F^{\mu\nu}\right) ,
\end{equation}
\end{widetext}
where $\phi$ is a neutral scalar, $V(\phi)$ is its potential, $Z(\phi)$ is the gauge kinetic function, $F_{\mu\nu} = \partial_\mu A_\nu - \partial_\nu A_\mu$ is the $U(1)$ field strength. This framework describes Einstein gravity coupled to a $U(1)$ gauge field and a scalar field, and is the standard arena for holographic systems at finite charge density~\cite{Faulkner:2009wj}. The black hole metric used throughout this paper is not chosen arbitrarily; it is a solution to the equations of motion derived from~\eqref{eq:EMS-action} that admits a smooth extremal limit, with the near-horizon geometry in the regime $T\to 0$ determined to leading order by the near-horizon dynamics.

The black hole solutions of interest take the planar form
\begin{equation}\label{eq:metric-planar}
    ds^2 = -f(r)\,dt^2 + \frac{dr^2}{f(r)} + h(r)\,d\vec{x}^2,
\end{equation}
where $\dd{\vec{x}}^{2}=dx_1^2+\cdots+dx_{d-1}^2$, $f(r_h) = 0$ defines the horizon radius, $h(r)$ controls the transverse geometry, and the Hawking temperature is $T = f'(r_h)/(4\pi)$. In models admitting two nearby horizons, we denote the outer and inner horizon radii by $r_+$ and $r_-$ respectively; the temperature is then controlled by the horizon separation and typically scales as $T\sim (r_+-r_-)$, with the near-extremal regime corresponding to $r_+\to r_-\equiv r_0$.

In the near-extremal regime, expanding $f(r)$ about the outer horizon $r_h = r_+$ with $f(r_+)=0$ and $f'(r_+)=4\pi T$, one obtains
\begin{equation}\label{eq:near-extremal-f}
    f(r) \approx 4\pi T(r - r_+) + \tfrac{1}{2}f''(r_+)(r - r_+)^2,
\end{equation}
where the linear term, proportional to $T$, vanishes in the strict extremal limit $T=0$, i.e.\ $r_+\to r_-$, leaving only the double zero.
Setting $\rho = r - r_+$, the near-horizon metric becomes
\begin{align}\label{eq:near-extremal-metric}
    ds^2 \approx &-\rho\!\left(4\pi T + \tfrac{1}{2}f''(r_+)\,\rho\right)dt^2
    + \frac{d\rho^2}{\rho\!\left(4\pi T + \tfrac{1}{2}f''(r_+)\,\rho\right)} \nonumber \\
    &+ h(r_+)\,d\vec{x}^2,
\end{align}
Equivalently, since $f(r)$ possesses simple zeros at both $r_+$ and $r_-$, the metric function admits the factored form
\begin{equation}\label{eq:near-extremal-f-factored}
    f(r) \simeq \frac{(r - r_+)(r - r_-)}{L_2^2},
\end{equation}
where $L_2^2 = 2/f''(r_0)$ is the AdS$_2$ radius introduced in Eq.~\eqref{eq:AdS2-radius-early}.
Substituting into Eq.~\eqref{eq:metric-planar} yields the manifestly two-horizon form
\begin{equation}\label{eq:near-extremal-metric-symmetric}
    ds^2 \simeq -\frac{(r - r_+)(r - r_-)}{L_2^2}\,dt^2
    + \frac{L_2^2\,dr^2}{(r - r_+)(r - r_-)}
    + h(r_0)\,d\vec{x}^2,
\end{equation}
which is the standard representation in the near-extremal literature~\cite{Faulkner:2009wj} and makes the emergent AdS$_2$ throat explicit upon taking $r_+ \to r_- \equiv r_0$.
In the strict extremal limit $T=0$, the linear term vanishes. Comparing the resulting metric with the standard AdS$_2$ metric in Poincaré coordinates,
\begin{equation}
    ds^2_{\mathrm{AdS}_2} = L_2^2\!\left(-\zeta^2 d\tau^2 + \frac{d\zeta^2}{\zeta^2}\right),
\end{equation}
one identifies the AdS$_2$ radius~\cite{Faulkner:2009wj} and finds that the near-horizon geometry approximately factorizes into $\mathrm{AdS}_2 \times \mathbb{R}^{d-1}$ with
\begin{equation}\label{eq:AdS2-radius-early}
    L_2^2 = \frac{2}{f''(r_h)}.
\end{equation}
This $\mathrm{AdS}_2$ throat is a robust feature of Einstein gravity: provided the extremal solution carries nonzero entropy and no symmetry-breaking instability intervenes at low temperature, it emerges in any extremal solution of Eq.~\eqref{eq:EMS-action} regardless of the specific form of $V(\phi)$ or $Z(\phi)$, and its appearance is a direct consequence of the double zero in $f(r)$ rather than a property of any particular model~\cite{Faulkner:2009wj}. In the near-extremal regime the linear term in Eq.~\eqref{eq:near-extremal-f} introduces $\mathcal{O}(T)$ corrections to this geometry; the AdS$_2$ structure is therefore approximate for $T>0$, becoming exact only at $T=0$.

The $\mathrm{AdS}_2$ factor carries its own infrared conformal symmetry, distinct from the asymptotic $\mathrm{AdS}_{d+1}$ structure. In the $\mathrm{AdS}_2/\mathrm{CFT}_1$ correspondence~\cite{Sachdev:2010um}, this geometry is the holographic dual of a zero-temperature quantum critical point, and low-energy excitations are organized into a tower of operators with IR scaling dimensions $\Delta_{\mathrm{IR}}$. A scalar field of mass $m$ with transverse momentum $k$ propagating in the emergent approximately AdS$_2$ near-horizon throat acquires an effective mass~\cite{Faulkner:2009wj}
\begin{equation}\label{eq:meff-early}
    m_{\mathrm{eff}}^2 = m^2 + \frac{k^2}{h(r_h)},
\end{equation}
and the corresponding IR conformal dimension satisfies
\begin{equation}\label{eq:IR-dim-early}
    \Delta_{\mathrm{IR}}(\Delta_{\mathrm{IR}} - 1) = m_{\mathrm{eff}}^2 \, L_2^2.
\end{equation}

This framework motivates a concrete physical expectation: if high-order pole-skipping is governed by the near-horizon approximately AdS$_2$ geometry rather than the full bulk spacetime, the pole-skipping momenta should be selected by the condition that the scalar field matches an integer IR conformal dimension in the throat. The systematic near-extremal expansion developed below confirms this expectation.

\subsection{Standard Near-Horizon Analysis}\label{sec:Standard Near-Horizon Analysis}

We begin by reviewing the standard near-horizon method for determining pole-skipping points~\cite{Blake:2017ris,Ahn:2020bks,Liu:2018kfw}, which motivates our systematic near-extremal approach.

We work with the $(d+1)$-dimensional metric Eq.~\eqref{eq:metric-planar}.
Introducing ingoing Eddington--Finkelstein coordinates $v=t+r_*$ with tortoise coordinate $\dd r_*=\dd r/f(r)$, the metric becomes
\begin{equation}
   \label{eq:EF-metric}
   ds^2=-f(r)dv^2+2dvdr+ h(r)\dd{\vec{x}}^2 .
\end{equation}

The Klein--Gordon equation for a scalar field $\Phi$ with mass $m$ is
\begin{equation}
\label{eq:KG}
\frac{1}{\sqrt{-g}}\partial_\mu(\sqrt{-g}g^{\mu\nu}\partial_\nu\Phi)-m^2\Phi=0\,.
\end{equation}
For plane wave fluctuations $\Phi(v,r,x_1)=e^{-i \omega v+ikx_1}\Phi(r)$,
the scalar field equation of motion Eq.~\eqref{eq:KG} reduces to
\begin{equation}\label{eq:scalar-EOM}
    \mathcal{S}=a(r)\Phi''(r)+b(r)\Phi'(r)+c(r)\Phi(r)=0,
\end{equation}
where the coefficients are
\begin{align}
	a(r)&=f(r)h(r),\\
	b(r)&=\frac{d-1}{2}f(r) h'(r)+h(r) \left(f'(r)-2 i \omega \right),\\
	c(r)&=-\left(i \omega \frac{d-1}{2} h'(r)+m^2 h(r)+k^2\right).
\end{align}

Expanding around the horizon $r=r_h$ with $f(r_h)=0$, we assume
\begin{align}
	\Phi(r)&=(r-r_h)^\lambda \sum_{i=0}^{\infty} \Phi_{i}\left(r-r_{h}\right)^{i}.\nonumber \\
    \mathcal{S}&=(r-r_h)^{\lambda-1} \sum_{i=0}^{\infty} S_{i}\left(r-r_{h}\right)^{i}.
\end{align}
Substituting into Eq.~\eqref{eq:scalar-EOM} and examining the leading order, the indicial equation yields two exponents
\begin{equation}
\lambda_1 = 0 , \qquad \lambda_2 = \frac{i\omega}{2\pi T},
\end{equation}
where $T = f'(r_h)/(4\pi)$ is the Hawking temperature. The exponent $\lambda_1$ corresponds to the ingoing mode, whereas $\lambda_2$ is generically outgoing. Imposing ingoing boundary conditions therefore selects $\lambda_1=0$.

Substituting $\lambda_1=0$ into the equation of motion order by order, we obtain the recursion relations
\begin{widetext}
\begin{equation}
   		\begin{aligned}
   &\mathcal{M}(\omega,k^2)\cdot \Phi\equiv
   \begin{pmatrix}
   M_{11} & (2\pi T-i\omega) & 0    & 0  &\cdots\\
   M_{21} & M_{22}& (4\pi T-i\omega)& 0   &\cdots\\
   M_{31} & M_{32}&  M_{33} &(6\pi T-i\omega) &\cdots\\
   	\vdots   &  \vdots&  \vdots  &\vdots   &\ddots\\
   \end{pmatrix}
   \times\begin{pmatrix}
   	\Phi_0\\
   	\Phi_1\\
   	\Phi_2 \\
   	\vdots \\
   \end{pmatrix}=0\,.
   	\end{aligned}
       \label{condition}
\end{equation}
\end{widetext}

At special points $(\omega_n, k_n^2)$ in the complex frequency-momentum plane, the recursion matrix admits a non-trivial null vector even with the ingoing boundary condition~\cite{Blake:2019otz}. These points are determined by
\begin{equation}\label{eq:ps-condition}
   \omega_{n}=-i2\pi Tn,\qquad \det\mathcal{M}^{(n)}(\omega_{n},k^2)=0\,.
\end{equation}
The $n$-th order condition involves an $n\times n$ determinant and yields a polynomial equation of degree $n$ in $k^2$, generically admitting $n$ distinct solutions $k_{n,q}^2$ with $q=1,\ldots,n$.

While systematic in principle, this method becomes increasingly intractable at high orders. The determinant involves $n$-th derivatives of the metric functions, and the algebraic complexity grows rapidly with $n$. This severely limits analytic control beyond the first few orders.

Historically, pole-skipping was discovered through its connection to quantum chaos: the leading pole-skipping point of the gravitational sound channel encodes the Lyapunov exponent $\lambda_L$ and butterfly velocity $v_B$ extracted from the out-of-time-ordered correlator (OTOC)~\cite{Grozdanov:2017ajz,Blake:2017ris},
  \begin{equation}\label{eq:chaos-relation}
      \omega = i\lambda_L, \qquad k = i\,\frac{\lambda_L}{v_B}.
  \end{equation}
While this relation is well established for the leading point, the physical interpretation of the higher-order pole-skipping points remains an open question. In the following, we develop a near-extremal expansion that provides analytic access to pole-skipping at arbitrary order and offers a new tool for investigating this structure.

\subsection{Systematic Near-Extremal Expansion Method}\label{sec:Systematic Near-Extremal Expansion Method}

The near-extremal limit $T\to 0$ introduces a natural hierarchy in the recursion matrix that fundamentally simplifies the pole-skipping structure.
Treating temperature as a small parameter,
\begin{equation}
	T = \frac{f'(r_h)}{4\pi} \to 0 ,
	\label{eq:HawkingT_near_extremal}
\end{equation}
the superdiagonal elements $(2j\pi T + i\omega_n) = 2\pi T(j-n)$ become parametrically small, while the diagonal elements retain $\mathcal{O}(T^0)$ contributions from the near-horizon geometry. This temperature-graded structure, rooted in the emergent AdS$_2$ geometry established in Sec.~\ref{sec:motivation}, serves as the organizing principle of our method. We now develop a systematic expansion that exploits this hierarchy to reduce the infinite-dimensional determinant problem to a finite algebraic equation.

The expansion is organized with $n$ held fixed while $T \to 0$, so that the Matsubara frequency $\omega_n = -i2\pi T n \to 0$. The resulting formulas are therefore reliable in the regime $nT \to 0$, i.e.\ for orders $n$ satisfying $n \ll (2\pi T)^{-1} \Lambda_{\mathrm{near\text{-}horizon}}$, where $\Lambda_{\mathrm{near\text{-}horizon}}$ denotes the relevant near-horizon energy scale. Corrections of order $nT$ enter at the next level of the expansion and can become important when $n$ grows parametrically large at fixed $T$; we comment on this regime in Sec.~\ref{sec:conclusion}.

Substituting $\omega_n = -i2\pi Tn$ into the diagonal elements $M_{jj}(\omega_n, k^2)$ yields
\begin{align}
	&M_{jj}(\omega_n, k^2)=\nonumber \\
     &-(k^2+ m^2 h(r_h) - \frac{j(j-1)}{2}h(r_h)f''(r_h))\nonumber \\ &
	+ \pi h'(r_h)(2 (-1 + j) (-3 + d + 2 j) - (-5 + d + 4 j) n)T.
\end{align}
When $h(r_h)$ and $f''(r_h)$ are not parametrically small,
the first three terms are $\mathcal{O}(T^0)$, while the last term is $\mathcal{O}(T)$.
This temperature-graded structure allows us to systematically expand the recursion relations in powers of $T$.

To obtain the $n$-th order pole-skipping solutions in this limit,
we decompose each matrix element according to its temperature dependence
\begin{equation}
	M_{ij} = M_{ij}^{(0)} +  M_{ij}^{(1)} \,T.
\end{equation}
This decomposition is exact: for the near-extremal backgrounds considered here, no terms beyond $\mathcal{O}(T)$ appear in the matrix elements.
The diagonal elements at $\omega = \omega_n$ are given by
\begin{align}
	&M_{jj}^{(0)} = -k^2- m^2 h(r_h) + \frac{j(j-1)}{2}h(r_h)f''(r_h), \quad \\
	&M_{jj}^{(1)} = \nonumber\\
    & \pi(2 (-1 + j) (-3 + d + 2 j) - (-5 + d + 4 j) n) h'(r_h),
\end{align}
The superdiagonal elements scale as $M_{j,j+1} = 2\pi T(j-n) \sim \mathcal{O}(T)$.

The temperature-graded structure of $\mathcal{M}^{(n)}$ evaluated at $\omega = \omega_n$ is therefore
\begin{equation}\label{eq:matrix-Tstructure}
    \mathcal{M}^{(n)}(\omega_n,k^2) =
    \begin{pmatrix}
    \mathcal{O}(T^0) & \mathcal{O}(T)   & 0               & \cdots \\
    \mathcal{O}(T^0) & \mathcal{O}(T^0) & \mathcal{O}(T)  & \cdots \\
    \mathcal{O}(T^0) & \mathcal{O}(T^0) & \mathcal{O}(T^0)& \cdots \\
    \vdots           & \vdots           & \vdots           & \ddots
    \end{pmatrix},
\end{equation}
where all entries on and below the diagonal ($i \geq j$) are $\mathcal{O}(T^0)$, superdiagonal entries ($j = i+1$) are $\mathcal{O}(T)$, and all entries two or more above the diagonal vanish identically. In special cases, individual entries may vanish or acquire higher-order $T$ corrections, but the generic scaling is as shown.

This structure directly implies the factorization of the determinant. By the Leibniz expansion $\det\mathcal{M}^{(n)} = \sum_{\sigma\in S_n}\mathrm{sgn}(\sigma)\prod_{i=1}^n M_{i,\sigma(i)}$, any permutation $\sigma \neq \mathrm{id}$ must include at least one element from strictly above the diagonal: the only permutation satisfying $\sigma(i)\leq i$ for all $i$ is the identity (since $\sigma(1)\geq 1$ forces $\sigma(1)=1$, then $\sigma(2)\neq 1$ and $\sigma(2)\leq 2$ forces $\sigma(2)=2$, and so on by induction). Since all entries two or more above the diagonal vanish, every $\sigma\neq\mathrm{id}$ must use at least one superdiagonal factor $\mathcal{O}(T)$. Therefore,
\begin{equation}\label{eq:det-leading}
    \det\mathcal{M}^{(n)}(\omega_n,k^2) = \prod_{\sigma=1}^n M_{\sigma\sigma}^{(0)} + \mathcal{O}(T)\,,
\end{equation}
and the pole-skipping condition $\det\mathcal{M}^{(n)}=0$ reduces at leading order to the extremal factorization condition,
\begin{equation}\label{eq:factorized}
    \prod_{\sigma=1}^n M_{\sigma\sigma}^{(0)} = 0\,.
\end{equation}
An alternative derivation via near-horizon recursion relations, which additionally establishes the coefficient scaling $\Phi_j \sim T^{-j}\Phi_0$, is presented in Appendix~\ref{app:iterative}.

This factorization is the key technical result of the near-extremal analysis. The $n$-th order pole-skipping condition, which at finite temperature requires evaluating an $n \times n$ determinant, reduces in the extremal limit to the extremal factorization condition, Eq.~\eqref{eq:factorized}. The physical origin of this simplification is the temperature-graded hierarchy~\eqref{eq:matrix-Tstructure}: by the Leibniz argument above, every non-identity permutation must use at least one superdiagonal factor $\mathcal{O}(T)$, so the determinant reduces to a product of diagonal elements at leading order in $T$. Each factor $M_{qq}^{(0)}$ depends on $k^2$, $m^2$, and the near-horizon geometric data $h(r_h)$ and $f''(r_h)$, but crucially does not involve the order $n$. Consequently, the momentum $k_{n,q}^2$ obtained by setting any single factor $M_{qq}^{(0)} = 0$ is entirely independent of $n$: all pole-skipping points sharing the same mode index $q$, regardless of their order, converge to the same momentum in the extremal limit. This $n$-independent degeneracy is a hallmark of the near-extremal regime and has no counterpart at finite temperature, where the full $n \times n$ determinant generically couples all matrix elements and produces $n$ distinct solutions that depend on $n$ in a complicated way.
When $M_{qq}^{(0)}=0$ for some $q\in\{1,\ldots,n\}$, the product~\eqref{eq:factorized} vanishes and a pole-skipping point arises with mode index $q$; the corresponding momenta are determined in Sec.~\ref{sec:Pole-Skipping Momenta in the Near-Extremal Limit}.

\subsection{Pole-Skipping Momenta in the Near-Extremal Limit}\label{sec:Pole-Skipping Momenta in the Near-Extremal Limit}

The extremal factorization condition, Eq.~\eqref{eq:factorized}, immediately yields the pole-skipping points in the extremal limit $T\to 0$:
\begin{align}\label{eq:extremal-momenta}
	&\omega_{n}=-i2\pi Tn,\nonumber\\
    & k_{n,q}^2  =- m^2 h(r_h) + \frac{q(q-1)}{2}h(r_h)f''(r_h) ,
\end{align}
where mode index $q=1,2,\ldots,n$. Note that since $\omega_n = -i2\pi Tn \to 0$ as $T\to 0$ at fixed $n$, the frequency in Eq.~\eqref{eq:extremal-momenta} collapses to zero in the strict extremal limit; the formula should be understood as the $T\to 0$ limit of the finite-temperature pole-skipping trajectory, labeled by the discrete index $n$ rather than by the frequency value.

Eq.~\eqref{eq:extremal-momenta} reveals a dramatic simplification compared to the standard determinant approach reviewed in Sec.~\ref{sec:Standard Near-Horizon Analysis}. At finite temperature, determining the $n$-th order pole-skipping points requires evaluating an $n\times n$ determinant involving $n$-th derivatives of metric functions, a calculation that rapidly becomes intractable. In the extremal limit, this complexity collapses: the pole-skipping momenta are given by a simple algebraic formula depending only on near-horizon data $h(r_h)$, $f''(r_h)$, and $m^2$. A key feature is that $k_{n,q}^2$ is independent of the order $n$ and the spacetime dimension $d$; the dimension only enters through higher-order temperature corrections. The $n$-th order yields $n$ discrete solutions labeled by the mode index $q$, each corresponding to a different solution of the extremal factorization condition, Eq.~\eqref{eq:factorized}. This reduction demonstrates that the temperature hierarchy identified in Sec.~\ref{sec:Systematic Near-Extremal Expansion Method} fundamentally reorganizes the pole-skipping structure, making pole-skipping points at large orders $n$ (with $nT \to 0$) analytically accessible.

\begin{figure}[htbp]
    \centering
    \includegraphics[width=0.5\textwidth]{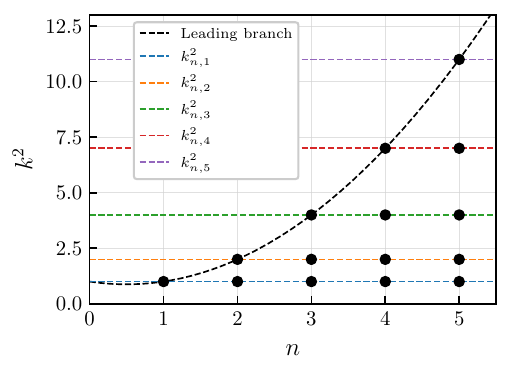}
    \caption{The first 5 orders of pole-skipping points predicted by Eq.~\eqref{eq:extremal-momenta}, where we set $m^2 = -1$, $h(r_h) = 1$, and $f''(r_h) = 1$. The black curves connect the leading pole-skipping branch ($q=n$) as a visual guide, corresponding to Eq.~\eqref{eq:max branch}; the colored dashed lines indicate identical momenta for modes with the same index $q$.}
    \label{fig:ps-pattern}
  \end{figure}

Taking the largest momentum $q=n$ as an example, we have
\begin{equation}\label{eq:max branch}
k_{n,n}^2 = - m^2 h(r_h) + \frac{n(n-1)}{2} h(r_h) f''(r_h).
\end{equation}
Since $\omega_n = -i\,2\pi Tn$ grows linearly with $n$ while $k_{n,q}^2$ depends only on $q$, pole-skipping points with the same mode index $q$ share the same momentum and form horizontal lines in the $(\omega, k^2)$ plane.
As shown in Fig.~\ref{fig:ps-pattern}, each column of points corresponds to a fixed order $n$.

  For the mode index $q=1$, we obtain
  \begin{equation}
  k_{n,1}^2 = - m^2 h(r_h).
  \end{equation}
  This mass term persists at all orders of the pole-skipping points, reflecting that the pole-skipping condition is already satisfied at the earliest stage of the near-horizon recursion relations.

  For $n \gg 1$ with $nT \to 0$, Eq.~\eqref{eq:max branch} gives the asymptotic growth
  \begin{equation}\label{eq:max branch asymptotic}
  k_{n,n} \approx \pm n \sqrt{-\frac{h(r_h) f''(r_h)}{2}},
  \end{equation}
  where the proportionality constant is determined entirely by the near-horizon geometry. When $h(r_h)f''(r_h)>0$, the square root is real and positive. This linear growth of $k_{n,n}$ with the discrete order $n$ is a direct consequence of the factorization structure: at each order $n$, the leading pole-skipping momentum is set by the diagonal element $M_{nn}^{(0)}=0$, which depends quadratically on $n$. Note that $n$ here is a discrete pole-skipping order label; the apparent linearity is a pattern among discrete points $(n, k_{n,n})$, not a continuous dispersion relation.

  The near-horizon geometric quantities appearing in Eq.~\eqref{eq:extremal-momenta} can be expressed in terms of black hole thermodynamic variables, providing a more physical interpretation of the pole-skipping momenta.

For a general $(d+1)$-dimensional black hole, the Hawking temperature and entropy density are given by~\cite{Wald:1999vt}
\begin{equation}
T = \frac{f'(r_h)}{4\pi}, \qquad s = \frac{h(r_h)^{(d-1)/2}}{4G_N},
\end{equation}
where $G_N$ is the Newton constant. The near-horizon geometry can thus be expressed as
\begin{equation}
h(r_h) = (4G_N s)^{2/(d-1)}, \qquad f''(r_h) = \frac{4\pi T}{c} \frac{\partial s}{\partial r_h},
\end{equation}
where $c = T(\partial s/\partial T)$ is the specific heat. In the extremal limit $T \to 0$, the finiteness of $f''(r_h)$ requires that the ratio $c/T$ approaches a finite limit. We define $\kappa_0 \equiv \lim_{T\to 0} c/T$, so that the pole-skipping momenta can be rewritten as
\begin{equation}\label{eq:f_double_prime_extremal}
    k_{n,q}^2 = \left(-m^2  + \frac{2 q(q-1)\pi }{\kappa_0} \frac{\partial s}{\partial r_h}\right)(4G_N s)^{2/(d-1)},
\end{equation}
which remains well-defined as $T \to 0$ provided $\kappa_0$ is finite and nonzero. This thermodynamic perspective clarifies why the pole-skipping structure simplifies at low temperature: the relevant physics is controlled by the interplay between entropy density $s$ and the extremal thermal response encoded in $\kappa_0$.

A striking consequence of Eq.~\eqref{eq:extremal-momenta} is that the pole-skipping momenta $k_{n,q}^2$ in the extremal limit depend on the mode index $q$ but not on the order $n$. At finite temperature, the $n$-th order pole-skipping condition yields $n$ generically independent momenta, so that up to order $n$ there are $n(n+1)/2$ distinct constraints on the retarded Green's function $G^R(\omega,k)$. In the extremal limit, this set collapses: pole-skipping points at all orders share the same discrete set of momenta $\{k_q^2\}_{q=1,2,\ldots}$, with each momentum appearing at every Matsubara frequency $\omega_n$ for which $n \geq q$. The analytic structure of $G^R$ at imaginary Matsubara frequencies is thus controlled by a single discrete spectrum determined entirely by near-horizon data. This collapse reflects the dominance of the infrared: in the low-temperature limit, the Green's function ``forgets'' the details of the bulk geometry and its pole-skipping skeleton is organized by the near-horizon region alone.

At the kinematic level, this independence has a transparent origin: since $\omega_n = -i2\pi Tn$, all Matsubara frequencies collapse to zero simultaneously as $T \to 0$, so that the frequency can no longer distinguish different orders $n$ and the remaining pole-skipping constraint is purely geometric. At finite temperature, the $n$-th order pole-skipping condition involves derivatives of the metric functions up to order $n$, coupling the pole-skipping spectrum to the bulk geometry at progressively larger radial distances beyond the horizon. In the extremal limit, the development of an infinitely deep AdS$_2$ throat effectively screens the asymptotic region: perturbations at all Matsubara frequencies are governed by the same near-horizon geometry, and the pole-skipping spectrum loses its sensitivity to the ultraviolet details of the bulk spacetime.

The pole-skipping structure in the extremal limit confirms the physical expectation established in Sec.~\ref{sec:Systematic Near-Extremal Expansion Method}. Substituting Eq.~\eqref{eq:extremal-momenta} into the effective AdS$_2$ mass Eq.~\eqref{eq:meff-early} and the IR dimension relation Eq.~\eqref{eq:IR-dim-early} gives
\begin{equation}\label{eq:DeltaIR-q}
    m_{\mathrm{eff}}^2 \, L_2^2 = q(q-1), \qquad \Delta_{\mathrm{IR}} = q.
\end{equation}
The mode index $q$ is thus identified with the IR conformal dimension in the emergent AdS$_2$ throat. Different pole-skipping modes at the same order correspond to excitations with distinct effective masses and hence distinct IR scaling dimensions $\Delta_{\mathrm{IR}} = 1, 2, \ldots, n$. The extremal factorization condition, Eq.~\eqref{eq:factorized}, selects precisely those momenta at which the scalar field matches an integer IR conformal dimension, confirming that the pole-skipping spectrum is organized by the AdS$_2$/CFT$_1$ conformal tower rather than the full bulk geometry.
This identification connects pole-skipping to the broader framework of emergent quantum criticality in holography~\cite{Faulkner:2009wj}. The discrete set of allowed pole-skipping momenta maps onto the conformal tower of the AdS$_2$/CFT$_1$ correspondence, with each value of $q$ corresponding to a distinct effective mass for the AdS$_2$ problem set by the parent-theory momentum $k$.

This $n$-independence admits a simple physical interpretation. The integer $n$ labels the Matsubara frequency $\omega_n = -i2\pi T n$, while $q$ labels the IR scaling dimension $\Delta_{\mathrm{IR}} = q$ in the emergent AdS$_2$ throat. In the extremal limit, all Matsubara frequencies collapse to zero, rendering the frequency label $n$ physically indistinguishable. The only remaining constraint is that the scalar field matches an integer IR conformal dimension $\Delta_{\mathrm{IR}} = q$ in the AdS$_2$ throat.

This degeneracy suggests that all pole-skipping points in a given branch $q$ are probing the same IR physics of the AdS$_2$ throat, regardless of how deep they are in the holographic frequency tower. Equivalently, a pole-skipping point at order $n$ with mode index $q$ probes the same IR scaling dimension $\Delta_{\mathrm{IR}} = q$ as the $q$-th order point, differing only by the frequency at which it is accessed. The degeneracy across $n$ therefore reflects the infrared dominance of the AdS$_2$ region, with the pole-skipping structure organized by the conformal tower of the emergent CFT$_1$ rather than by the frequency label.

The asymptotic slope of the leading pole-skipping branch at large $q = n$ can also be understood in this language. Using Eq.~\eqref{eq:AdS2-radius-early}, the proportionality constant becomes
\begin{equation}\label{eq:v-infty}
    \sqrt{\frac{h(r_h)\, f''(r_h)}{2}} = \frac{\sqrt{h(r_h)}}{L_2},
\end{equation}
which is the ratio of the transverse geometric scale $\sqrt{h(r_h)}$ to the AdS$_2$ radius. This ratio controls the rate at which the pole-skipping momenta grow with Matsubara frequency and reflects the relative depth of the near-horizon throat.

\subsection{Leading Temperature Corrections in the Near-Extremal Regime}\label{sec:leading-temp-corrections}

To obtain the leading correction away from extremality, we systematically include $\mathcal{O}(T)$ terms in the recursion relations.

Near a pole-skipping point where $M_{qq}^{(0)} = 0$, we write
\begin{equation}\label{eq:T-correction}
    k_{n,q}^2 = - m^2 h(r_h) + \frac{q(q-1)}{2}h(r_h)f''(r_h) + C_{n,q}\,T + \mathcal{O}(T^2),
    \end{equation}
where the correction coefficient is
\begin{equation}
C_{n,q} = -M_{qq}^{(1)} + \frac{M_{q+1,q}^{(0)} M_{q,q+1}^{(0)}}{M_{q+1,q+1}^{(0)}} + \frac{M_{q,q-1}^{(0)} M_{q-1,q}^{(0)}}{M_{q-1,q-1}^{(0)}}.
\end{equation}
Here the boundary terms vanish: the last term is absent when $q=1$ (no $j=0$ row), and the second term is absent when $q=n$ (no $j=n+1$ row). The detailed calculation can be found in Appendix~\ref{sec:app}. Using similar methods, higher-order corrections in $T$ can be obtained in a systematic fashion.
This coefficient is real, ensuring $k^2$ remains real in the near-extremal regime, which provides a consistency check of the expansion.
The linear-in-$T$ correction captures the leading effect of finite temperature on the near-horizon geometry, and the coefficient $C_{n,q}$ depends on both the order $n$ and the mode index $q$, reflecting how different pole-skipping modes respond differently to thermal perturbations.

\subsection{Consistency Check in Simple Holographic Backgrounds}\label{sec:comparison}

As a consistency check on our method, we verify that our near-extremal expansion, including higher-order temperature corrections, correctly reproduces the known first-order pole-skipping results in the BTZ and planar AdS-Schwarzschild backgrounds~\cite{Blake:2019otz}. Since these backgrounds lack a smooth extremal limit with an emergent AdS$_2$ throat (the extremal limit renders the horizon an irregular singular point, invalidating higher-order Frobenius expansions~\cite{Natsuume:2020snz}), only the first pole-skipping point is accessible. Our method reproduces the known results $\lambda_L = 2\pi T$ and $v_B^2 = d/2(d-1)$~\cite{Blake:2019otz} in perfect agreement. The higher-order near-extremal structure that is the focus of this work requires a smooth extremal limit; this is verified numerically using the Dyonic Gubser--Rocha model in Sec.~\ref{sec:GR}.

Our systematic near-extremal expansion reveals a hierarchical structure in the pole-skipping problem:

The matrix $\mathcal{M}$ exhibits a natural hierarchy where diagonal elements contain $\mathcal{O}(T^0)$ and $\mathcal{O}(T)$ contributions, while superdiagonals are $\mathcal{O}(T)$.
Near-horizon coefficients scale as $\Phi_j \sim T^{-j}\Phi_0$ for $j < n$, simplifying the determinant condition.
The $n$-th order pole-skipping condition factorizes into $n$ independent algebraic equations, each determining one solution.
In the extremal limit, pole-skipping is controlled entirely by near-horizon geometry and the scalar field mass. These geometric quantities admit a thermodynamic interpretation in terms of entropy density $s$, specific heat $c$, and the extremal specific heat scale $\kappa_0 = \lim_{T\to 0} c/T$.

This structure provides analytic access to pole-skipping points at large orders $n$ (in the regime $nT \to 0$) and clarifies the organizing principle of holographic Green's functions in the low-temperature regime.

In the following section, we verify these predictions numerically in the Dyonic Gubser--Rocha model.

\section{Pole-Skipping in the Dyonic Gubser--Rocha Model}\label{sec:GR}

\subsection{Dyonic Gubser--Rocha Action and Black Hole Geometry}

We apply the near-extremal pole-skipping analysis to the dyonic Gubser--Rocha model~\cite{Ge:2023yom,Xu:2023qlu,Ishigaki:2024djz,Gubser:2008yx,Donos:2014cya}, which provides a holographic realization of momentum relaxation and finite charge density. This model serves as a concrete testbed owing to its rich parameter space, encompassing magnetic field, charge density, and momentum relaxation, and its smooth extremal limit, providing a nontrivial check that the near-extremal pole-skipping structure is universal across a broad class of holographic systems. The action consists of Einstein gravity coupled to an axio-dilaton field and axion fields that break translational symmetry, as in Ref.~\cite{Ge:2023yom}:

\begin{align} 
    S &= \frac{1}{2 \kappa^2} \int \mathrm{d}^4 x\left(\mathcal{L}_1+\mathcal{L}_2\right),\\
    \frac{\mathcal{L}_1}{\sqrt{-g}} &= R-\frac{3}{2} \frac{\partial_\mu \tau \partial^\mu \bar{\tau}}{(\operatorname{Im} \tau)^2}-\frac{1}{4} e^{-\phi} F^2+\frac{1}{4} \chi F \tilde{F}+\frac{3}{L^2} \frac{\tau \bar{\tau}+1}{\operatorname{Im} \tau},\\
    \frac{\mathcal{L}_2}{\sqrt{-g}} &= -\frac{1}{2} \sum_{I=1,2}\left(\partial \psi_I\right)^2,
\end{align}
Here $\tau = \chi + i e^{-\phi}$ is the axio-dilaton, $\tilde{F}^{\mu\nu} = \frac{1}{2 \sqrt{-g}} \epsilon^{\mu\nu\rho\sigma} F_{\rho\sigma}$ is the dual field strength, $\kappa$ is the gravitational constant, and $L$ is the AdS radius. The Lagrangian $\mathcal{L}_1$ describes the S-completed Gubser-Rocha model, while $\mathcal{L}_2$ introduces momentum relaxation through axion fields. Expanding $\tau$ in $\mathcal{L}_1$, we obtain
\begin{align}
	\begin{split}
	\frac{\mathcal{L}_1}{\sqrt{-g}}
	=&
	R - \frac{3}{2} (\partial \phi)^2 - \frac{3}{2} e^{2\phi} (\partial\chi)^2 
	-\frac{1}{4} e^{-\phi} F^2 + \frac{1}{4} \chi F \tilde{F}  \\ & + \frac{1}{L^2}(6\cosh\phi + 3\chi^2 e^{\phi}).
\end{split}
\end{align}
The scalar masses are $m_{\chi}^2=m^2_{\phi} = - 2/L^2$. Under standard quantization, the bulk scalar field is dual to a boundary operator $\mathcal{O}$ with conformal dimension satisfying $\Delta(\Delta - d) = m^2 L^2$, yielding $\Delta_{-}=1$ and $\Delta_{+}=2$. Both masses lie in the range $-d^2/4 \leq m^2 L^2 \leq -d^2/4 +1$. We set $2\kappa^2 = L = 1$ hereafter.

The dyonic Gubser--Rocha black hole solution~\cite{Ge:2023yom,Xu:2023qlu,Ishigaki:2024djz} takes the form~\cite{Ge:2023yom}
 \begin{equation}
	\dd{s}^2 = - f(r) \dd{t}^2 + \frac{\dd{r}^2}{f(r)} + h(r)(\dd{x}^2 + \dd{y}^2),
\end{equation}
with
\begin{align}\label{eq:solution_1}
	f(r) &= h(r)\left(
		1 +\frac{n^2 + B^2}{3 \rho (r-\rho)^3} - \frac{\beta^2 }{2 (r-\rho)^2}
	\right), \nonumber \\
	h(r) &= \sqrt{r(r-\rho)^3}.
\end{align}
Here $n$ is the charge density, $B$ is an external magnetic field orthogonal to the $x$--$y$ plane, $\beta$ characterizes momentum relaxation, and $\rho$ determines the location of the curvature singularity. The horizon radius $r_h$ is the largest real root of $f(r_h)=0$. The parameter range is $\rho>0$ and $\beta > 0$.

Defining the effective charge density $n_{\mathrm{eff}}^2 \equiv n^2 + B^2$, the blackening factor simplifies to
\begin{equation}
	f(r) = h(r)\left(
		1 + \frac{n_{\mathrm{eff}}^2}{3 \rho (r-\rho)^3} - \frac{\beta^2 }{2 (r-\rho)^2}
	\right) .
\end{equation}
The solution is thus characterized by $(\beta,\rho,n_{\mathrm{eff}})$. The horizon radius satisfies
\begin{equation}
  (r_h-\rho)^3  - \frac{\beta^2}{2} (r_h-\rho)+\frac{n_{ \mathrm{eff}}^2}{3 \rho} = 0 ,
\end{equation}
with explicit solution
\begin{equation}
r_{h}
= \rho + \sqrt{\frac{2}{3}}\,\beta\,
\cos\!\left[
\frac{1}{3}\arccos\!\left(
-\frac{\sqrt{6}\,n_{\mathrm{eff}}^2}{\beta^3\,\rho}
\right)
\right].
\end{equation}
The Hawking temperature is
\begin{equation}
	T = \frac{r_h}{8\pi\sqrt{r_h(r_h - \rho)^3}}\left(
		6(r_h-\rho)^2 - \beta^2 \right),
\end{equation}
and the near-extremal limit is approached as $\rho \beta^3-\sqrt{6}n_{\mathrm{eff}}^2 \to 0^+$.

In the near-extremal regime $\rho\beta^3 - \sqrt{6}n_{\mathrm{eff}}^2 \ll \rho\beta^3$,
the temperature $T$ is small but nonzero, and the outer horizon $r_+$ approaches
the extremal-limit horizon location
\begin{equation}\label{eq:r0-DGR}
    r_0 = \rho + \frac{\beta}{\sqrt{6}},
\end{equation}
where the double-zero condition $f(r_0)=f'(r_0)=0$ holds only in the strict
extremal limit $\rho\beta^3 = \sqrt{6}n_{\mathrm{eff}}^2$.
At finite $T$, the actual horizon satisfies $f'(r_+)=4\pi T\neq 0$,
with $r_+\to r_0$ as $T\to 0$.

Following the near-extremal expansion of Sec.~\ref{sec:motivation},
setting $\xi = r-r_+$, the near-horizon metric takes the
form~\eqref{eq:near-extremal-metric}
\begin{align}\label{eq:near-extremal-metric-DGR}
    \dd{s}^2 \approx{} &
    -\xi\!\left(4\pi T+\tfrac{1}{2}f''(r_+)\,\xi\right)\dd{t}^2
    +\frac{\dd{\xi}^2}{\xi\!\left(4\pi T+\tfrac{1}{2}f''(r_+)\,\xi\right)}
    \nonumber\\
    &+h(r_+)\!\left(\dd{x}^2+\dd{y}^2\right),
\end{align}
where $f'(r_+)=4\pi T$ encodes the simple zero at the horizon, marking the
finite-temperature deviation from the extremal double-zero structure.
This near-horizon geometry describes a finite-temperature AdS$_2$ black
hole throat, whose zero-temperature near-horizon limit becomes
AdS$_2\times\mathbb{R}^2$.

Since $r_+-r_0=\mathcal{O}(T)$, a Taylor expansion gives
\begin{equation}
    f''(r_+) = f''(r_0)+\mathcal{O}(r_+-r_0) = f''(r_0)+\mathcal{O}(T);
\end{equation}
to leading order in $T$ it is therefore determined by the extremal
near-horizon data,
\begin{equation}\label{eq:fpp-DGR}
    f''(r_0) = \frac{36}{\beta^2}\,h(r_0),\qquad
    h(r_0) = \sqrt{r_0\,(r_0-\rho)^3}.
\end{equation}
The AdS$_2$ radius $L_2^2=2/f''(r_0)$ Eq.~\eqref{eq:AdS2-radius-early},
derived from the extremal near-horizon expansion, controls the IR
geometric scale of the near-extremal black hole throat.
As $T\to 0$, the finite-temperature pole-skipping trajectories
$k_{n,q}(T)$ converge to the discrete limiting values $k_q$ ,the
$T\to 0$ endpoints of the near-extremal trajectories determined by
$L_2^2$ and $h(r_0)$ via Eq.~\eqref{eq:extremal-momenta}.

\subsection{Numerical Results for Pole-Skipping Momenta}

We now validate the analytic framework of Sec.~\ref{sec:near-extremal} against numerical solutions of the full determinant condition Eq.~\eqref{eq:ps-condition} in the dyonic Gubser--Rocha model. The verification proceeds in three stages: (i) convergence of numerical pole-skipping momenta to the extremal-limit formula Eq.~\eqref{eq:extremal-momenta} as $T\to 0$, (ii) universality of the pole-skipping trajectories across different orders $n$, and (iii) quantitative agreement of the leading temperature correction with Eq.~\eqref{eq:T-correction}.

We fix $\beta = 1$ and vary $n_{\mathrm{eff}}$ and $\rho$ to explore the near-extremal regime. For the extremal configuration $\rho = 1$ and $n_{\mathrm{eff}}^2 = \frac{1}{\sqrt{6}}$, the near-horizon data are $h(r_h) = \frac{\sqrt{1+\sqrt{6}}}{6}$ and $f''(r_h) = 6 \sqrt{1+\sqrt{6}}$. Eq.~\eqref{eq:extremal-momenta} then predicts the extremal pole-skipping momenta listed in Table~\ref{tab:extremal-momenta}, which are independent of the order $n$.

\sisetup{
table-number-alignment=center,
round-mode=places,
round-precision=6
}

\begin{table}[t]
\centering
\begin{tabular}{c S[table-format=2.6]}
\toprule
{$q$} & {$k_{n,q}^2$} \\
\midrule
1 & 0.6190934001320154 \\
2 & 4.068583142915194 \\
3 & 10.967562628481549 \\
4 & 21.316031856831085 \\
5 & 35.11399082796379 \\
6 & 52.36143954187968 \\
\bottomrule
\end{tabular}
\caption{Extremal pole-skipping momenta $k_{n,q}^2$ for different mode indices $q$, computed from Eq.~\eqref{eq:extremal-momenta} with $\rho=1$, $n_{\mathrm{eff}}^2=1/\sqrt{6}$, and $\beta=1$.}
\label{tab:extremal-momenta}
\end{table}

\begin{figure*}[htbp]
    \centering
    \begin{subfigure}{0.3\textwidth}
        \centering
        \includegraphics[width=\textwidth]{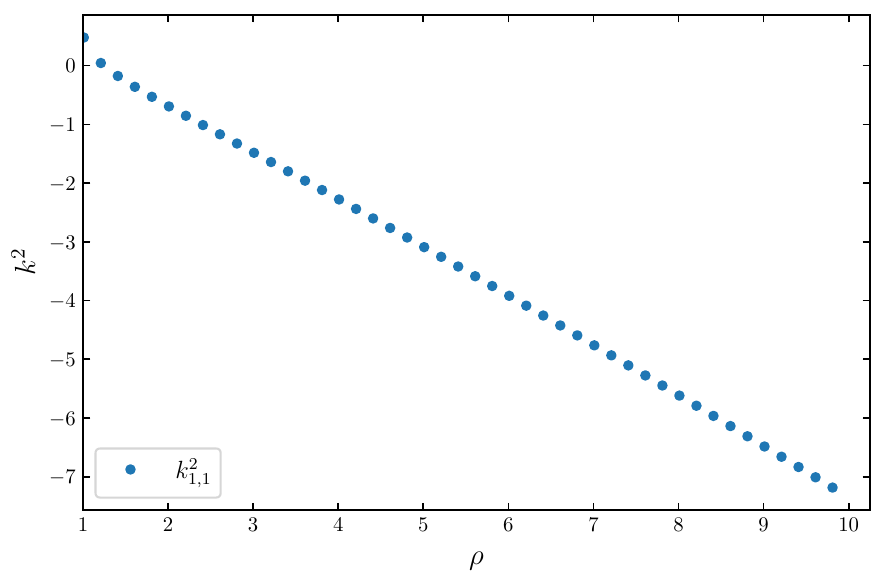}
        \caption{$n=1$}
        \label{}
    \end{subfigure}
    \begin{subfigure}{0.3\textwidth}
        \centering
        \includegraphics[width=\textwidth]{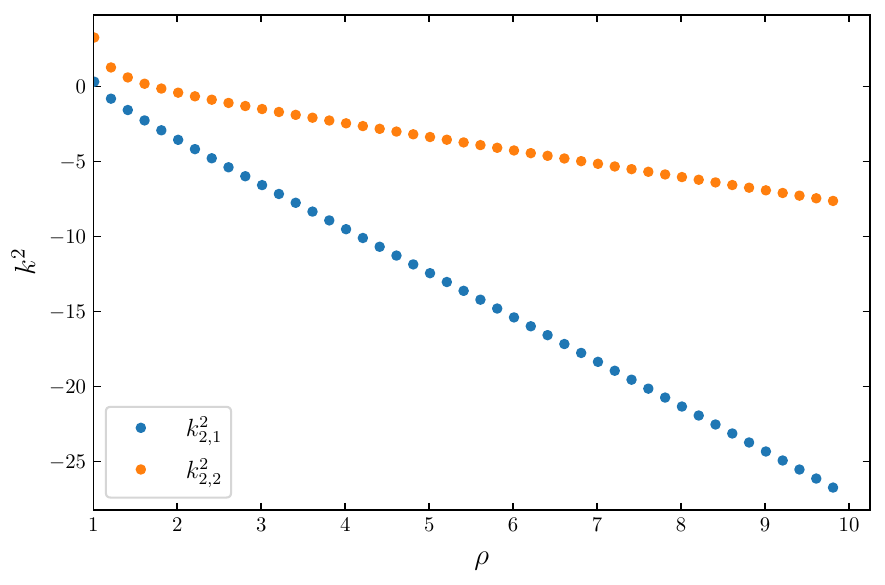}
        \caption{$n=2$}
        \label{}
    \end{subfigure}
    \begin{subfigure}{0.3\textwidth}
        \centering
        \includegraphics[width=\textwidth]{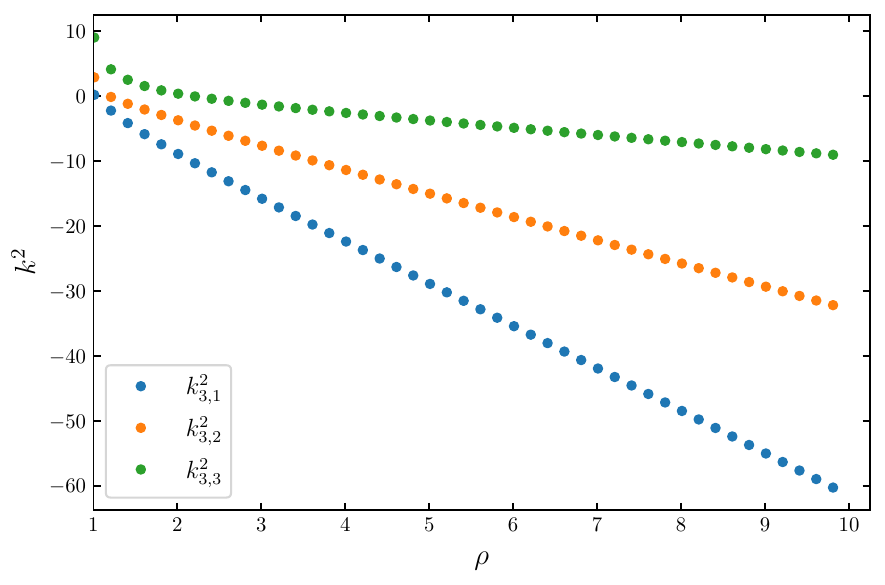}
        \caption{$n=3$}
        \label{}
    \end{subfigure}
    \begin{subfigure}{0.3\textwidth}
        \centering
        \includegraphics[width=\textwidth]{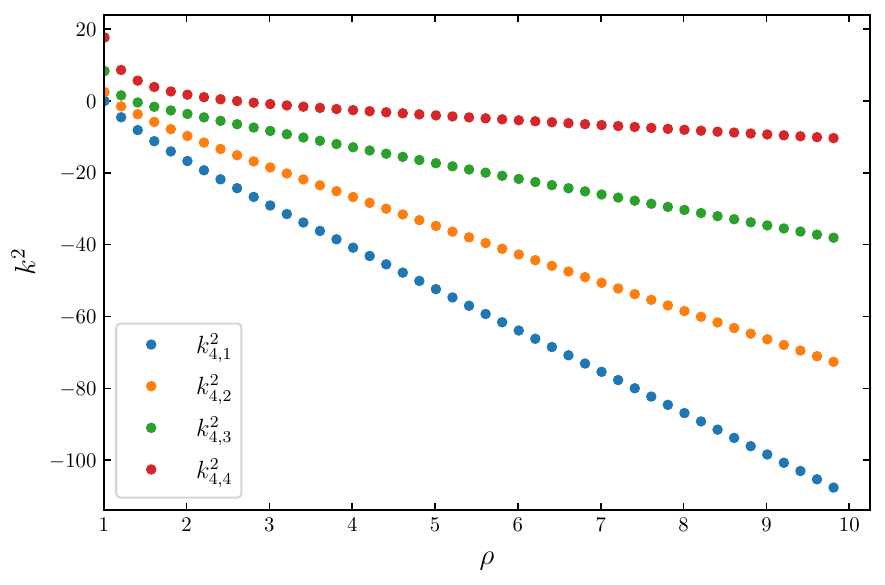}
        \caption{$n=4$}
        \label{}
    \end{subfigure}
    \begin{subfigure}{0.3\textwidth}
        \centering
        \includegraphics[width=\textwidth]{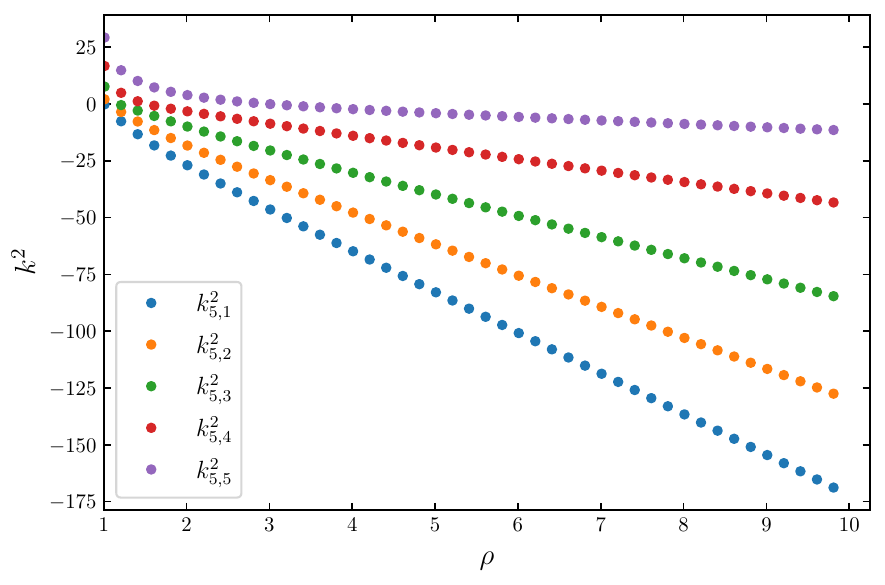}
        \caption{$n=5$}
        \label{}
    \end{subfigure}
    \begin{subfigure}{0.3\textwidth}
        \centering
        \includegraphics[width=\textwidth]{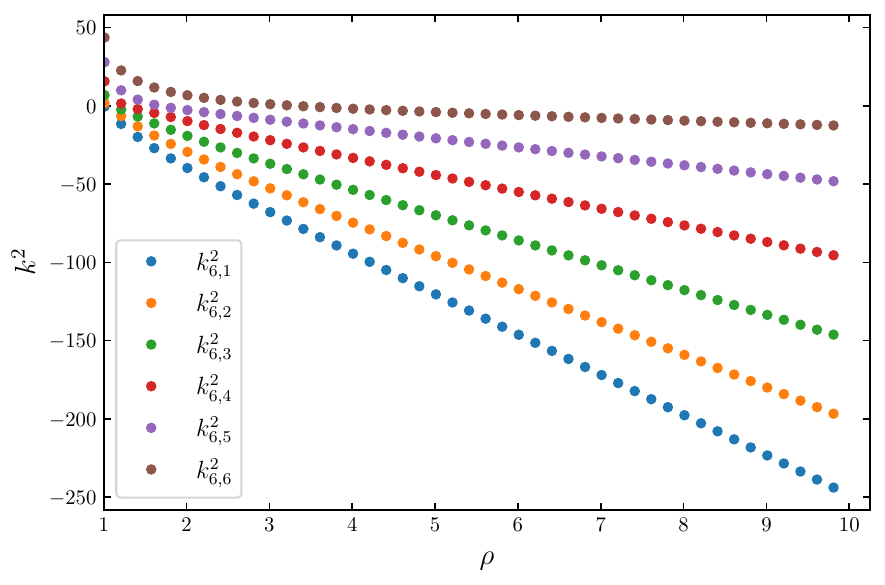}
        \caption{$n=6$}
        \label{}
    \end{subfigure}
 
    \caption{Pole-skipping momenta $k^2$ as a function of $\rho$ for orders $n=1$ through $n=6$, with $\beta = 1$ and $n_{\mathrm{eff}}^2 = 1/\sqrt{6}$. Different colors correspond to different mode indices $q$.}
    \label{fig:k2-rho}
\end{figure*}

Fig.~\ref{fig:k2-rho} shows the dependence of pole-skipping momenta on $\rho$ at fixed $n_{\mathrm{eff}}^2 = \frac{1}{\sqrt{6}}$, obtained by numerically solving Eq.~\eqref{eq:ps-condition}. As $\rho \to 1$, $k^2$ converges to the values in Table~\ref{tab:extremal-momenta} for all orders $n$ simultaneously, confirming the analytic prediction. At large $\rho$, $k^2$ exhibits approximately linear growth in $\rho$, consistent with the generic high-temperature behavior expected for holographic pole-skipping.

\begin{figure*}[htbp]
    \centering
    \begin{subfigure}{0.3\textwidth}
        \centering
        \includegraphics[width=\textwidth]{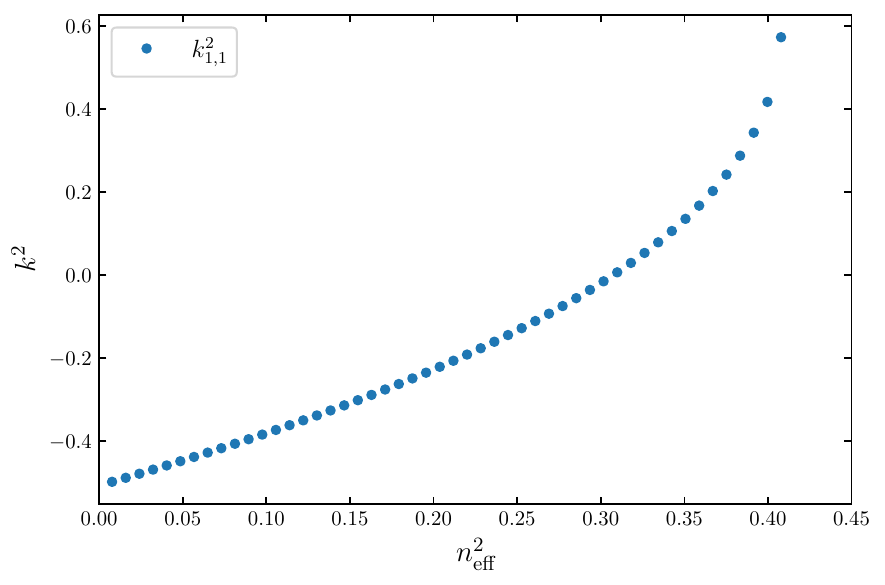}
        \caption{$n=1$}
        \label{}
    \end{subfigure}
    \begin{subfigure}{0.3\textwidth}
        \centering
        \includegraphics[width=\textwidth]{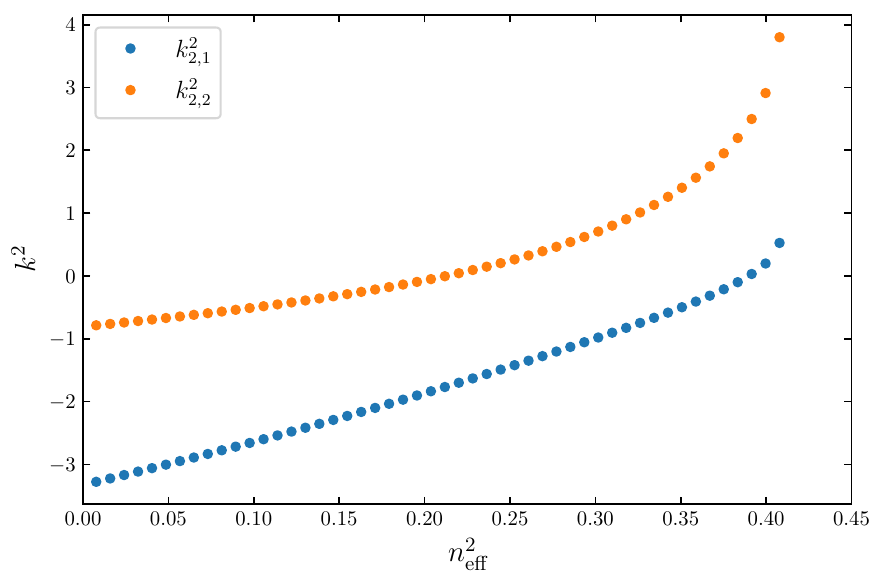}
        \caption{$n=2$}
        \label{}
    \end{subfigure}
    \begin{subfigure}{0.3\textwidth}
        \centering
        \includegraphics[width=\textwidth]{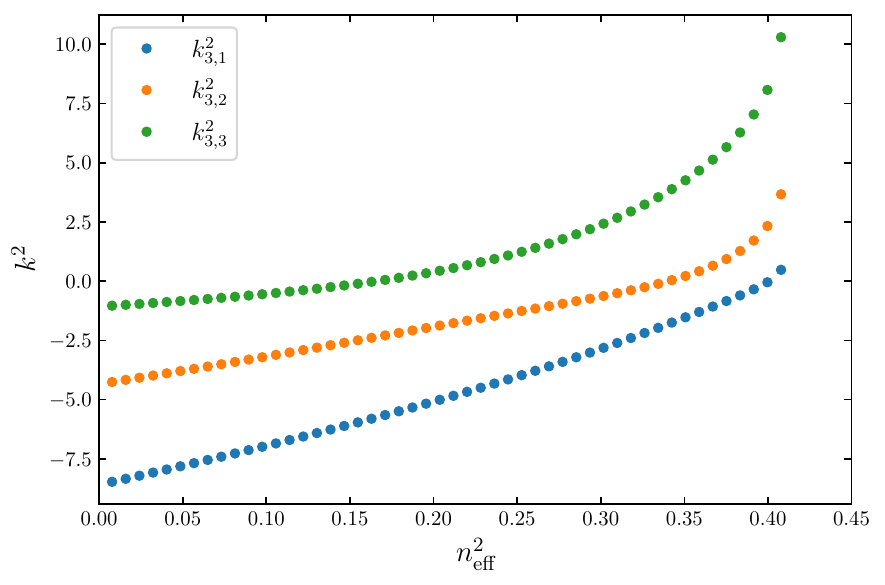}
        \caption{$n=3$}
        \label{}
    \end{subfigure}
    \begin{subfigure}{0.3\textwidth}
        \centering
        \includegraphics[width=\textwidth]{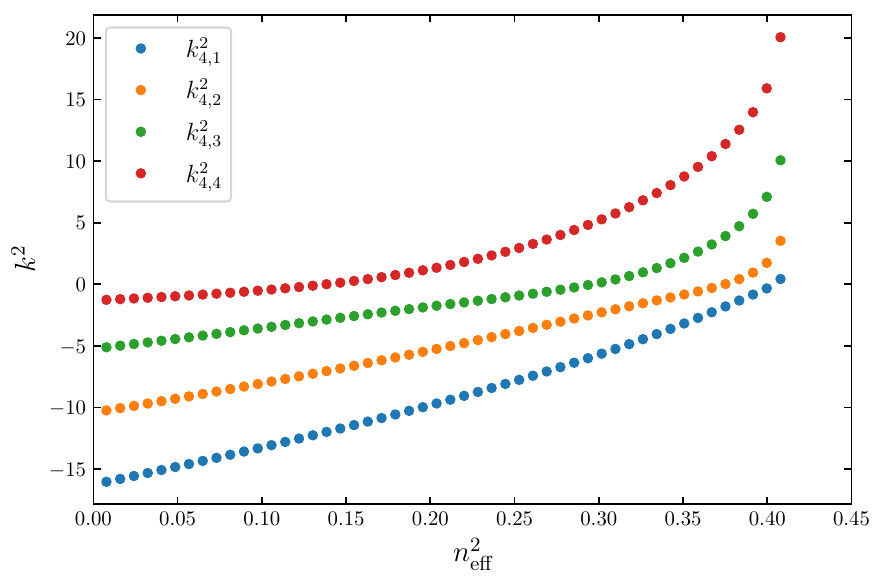}
        \caption{$n=4$}
        \label{}
    \end{subfigure}
    \begin{subfigure}{0.3\textwidth}
        \centering
        \includegraphics[width=\textwidth]{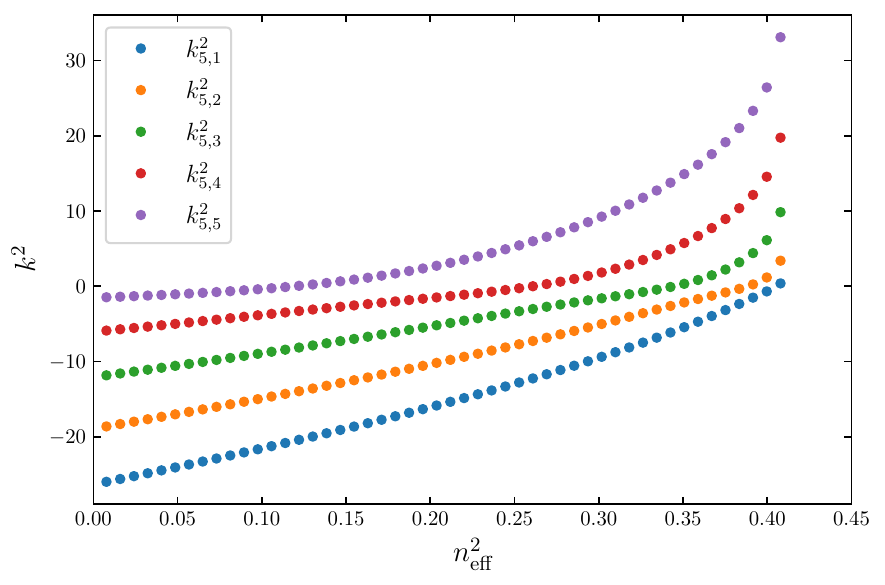}
        \caption{$n=5$}
        \label{}
    \end{subfigure}
    \begin{subfigure}{0.3\textwidth}
        \centering
        \includegraphics[width=\textwidth]{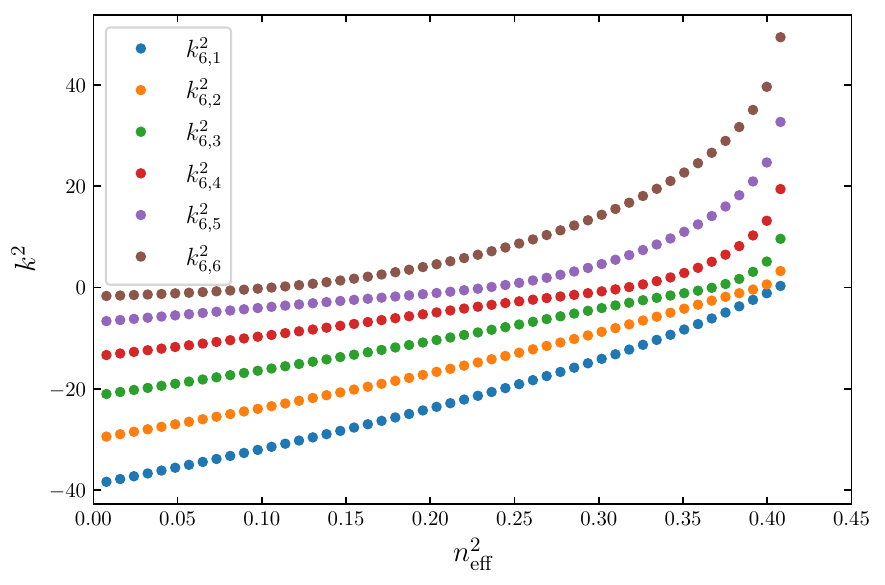}
        \caption{$n=6$}
        \label{}
    \end{subfigure}
 
    \caption{Pole-skipping momenta $k^2$ as a function of $n_{\mathrm{eff}}^2$ for orders $n=1$ through $n=6$, with $\beta = 1$ and $\rho = 1$. Different colors correspond to different mode indices $q$.}
    \label{fig:k2-neff}
\end{figure*}

Fig.~\ref{fig:k2-neff} shows the dependence on $n_{\mathrm{eff}}$ at fixed $\rho = 1$. The extremal limit $n_{\mathrm{eff}}^2 \to \frac{1}{\sqrt{6}}$ again reproduces the values in Table~\ref{tab:extremal-momenta}. Together, Figs.~\ref{fig:k2-rho} and~\ref{fig:k2-neff} demonstrate that the extremal momenta are approached robustly along two independent paths in parameter space, confirming that Eq.~\eqref{eq:extremal-momenta} captures the universal near-horizon structure rather than an artifact of a particular parameter choice.

\begin{figure*}[htbp]
    \centering
    \begin{subfigure}{0.3\textwidth}
        \centering
        \includegraphics[width=\textwidth]{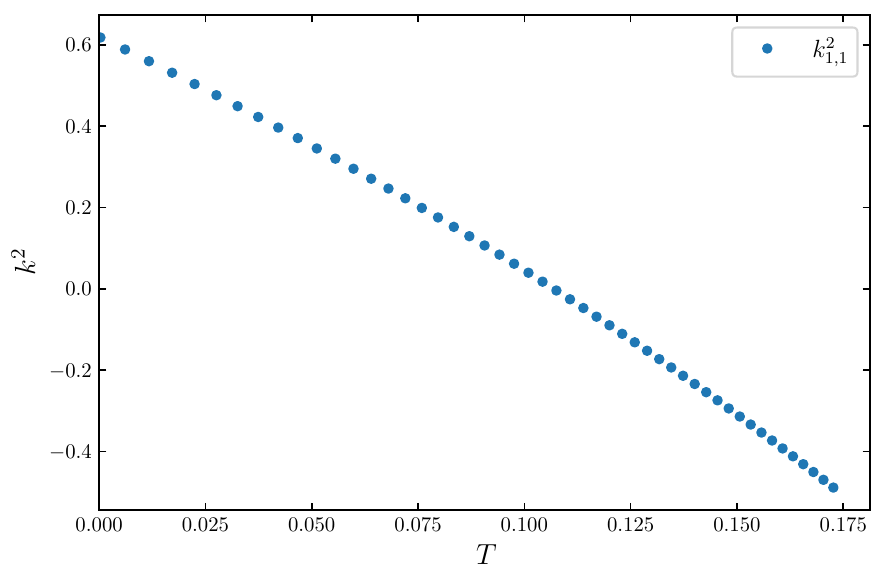}
        \caption{$n=1$}
        \label{}
    \end{subfigure}
    \begin{subfigure}{0.3\textwidth}
        \centering
        \includegraphics[width=\textwidth]{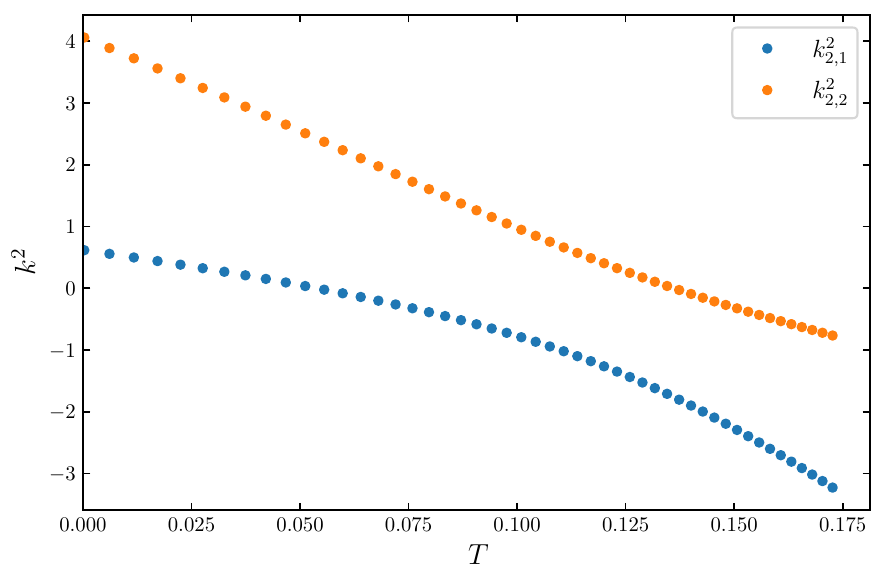}
        \caption{$n=2$}
        \label{}
    \end{subfigure}
    \begin{subfigure}{0.3\textwidth}
        \centering
        \includegraphics[width=\textwidth]{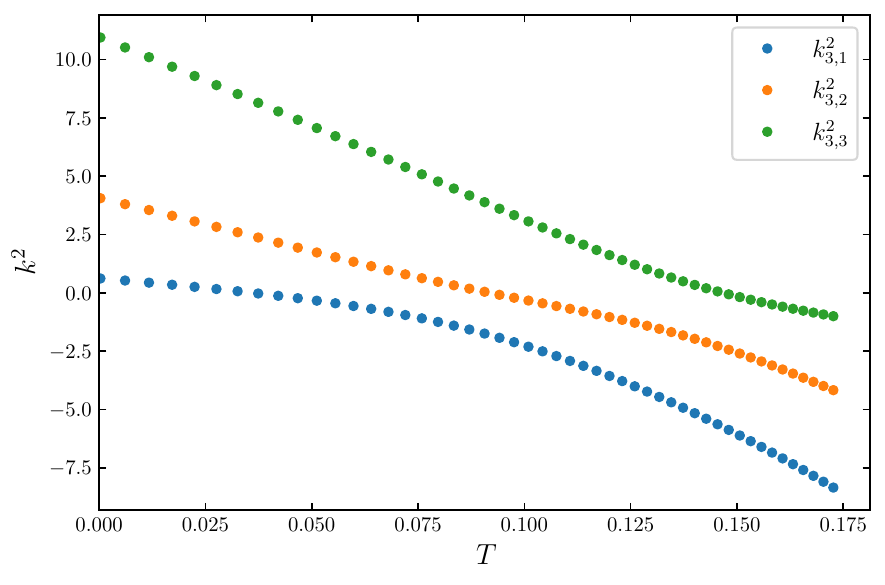}
        \caption{$n=3$}
        \label{}
    \end{subfigure}
    \begin{subfigure}{0.3\textwidth}
        \centering
        \includegraphics[width=\textwidth]{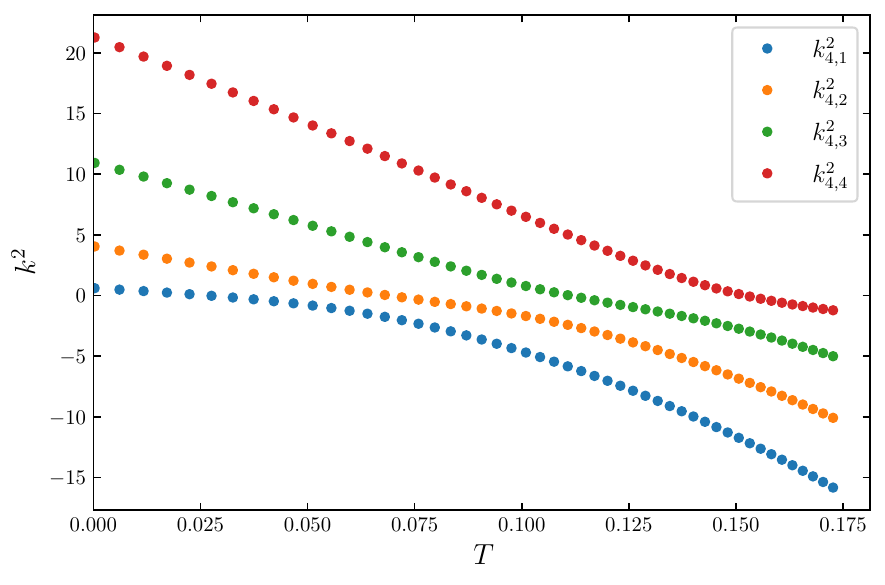}
        \caption{$n=4$}
        \label{}
    \end{subfigure}
    \begin{subfigure}{0.3\textwidth}
        \centering
        \includegraphics[width=\textwidth]{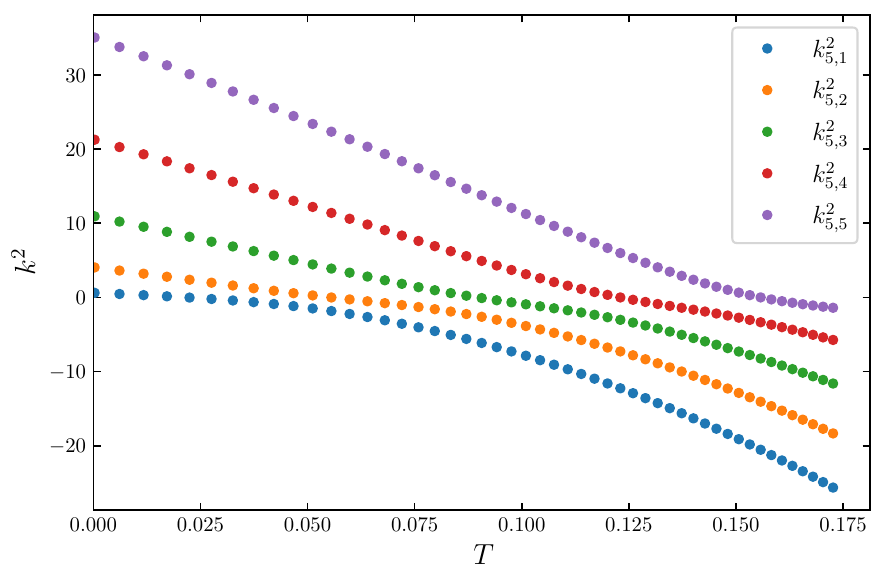}
        \caption{$n=5$}
        \label{}
    \end{subfigure}
    \begin{subfigure}{0.3\textwidth}
        \centering
        \includegraphics[width=\textwidth]{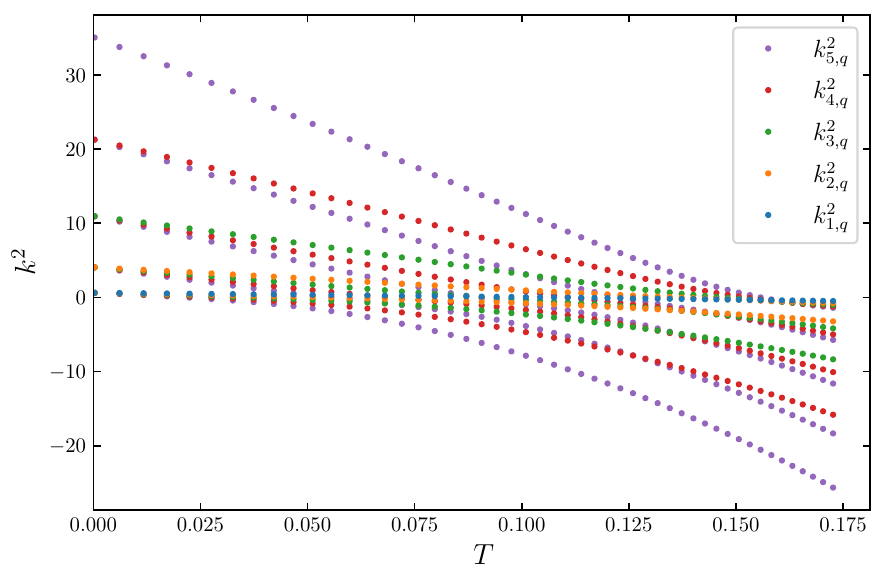}
        \caption{$n=1,2,3,4,5$}
        \label{}
    \end{subfigure}

    \caption{Temperature dependence of pole-skipping momenta $k^2(T)$ for orders $n=1$ through $n=5$. Panels (a)--(e) show individual orders; panel (f) overlays all curves.}
    \label{fig:k2-T}
\end{figure*}

We now fix $\rho = 1$ and vary $n_{\mathrm{eff}}$ to control the temperature. Fig.~\ref{fig:k2-T} displays $k^2$ as a function of $T$ for orders $n=1$ through $n=5$. As $T \to 0$, all branches converge to the extremal values in Table~\ref{tab:extremal-momenta}, which are independent of $n$ and determined solely by the mode index $q$ via Eq.~\eqref{eq:extremal-momenta}. The ordering of branches at fixed $n$ where larger $q$ corresponds to larger $k^2$ directly reflects the mode-index dependence of the extremal momenta. The convergence to these values is approximately linear in $T$, indicating that the leading temperature correction is first-order, consistent with Eq.~\eqref{eq:T-correction}. We examine this temperature dependence in detail below.

\begin{figure*}[htbp]
    \centering
    \begin{subfigure}{0.3\textwidth}
        \centering
        \includegraphics[width=\textwidth]{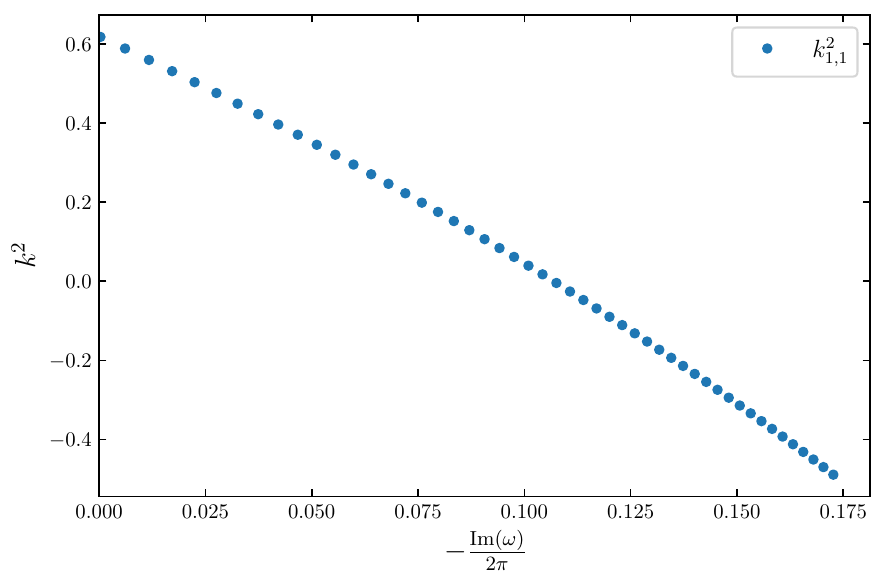}
        \caption{$n=1$}
        \label{}
    \end{subfigure}
    \begin{subfigure}{0.3\textwidth}
        \centering
        \includegraphics[width=\textwidth]{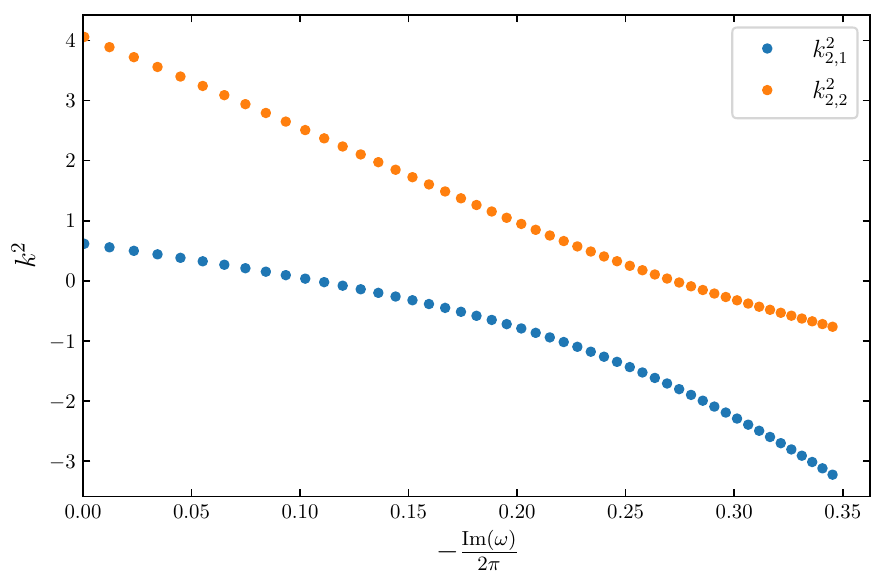}
        \caption{$n=2$}
        \label{}
    \end{subfigure}
    \begin{subfigure}{0.3\textwidth}
        \centering
        \includegraphics[width=\textwidth]{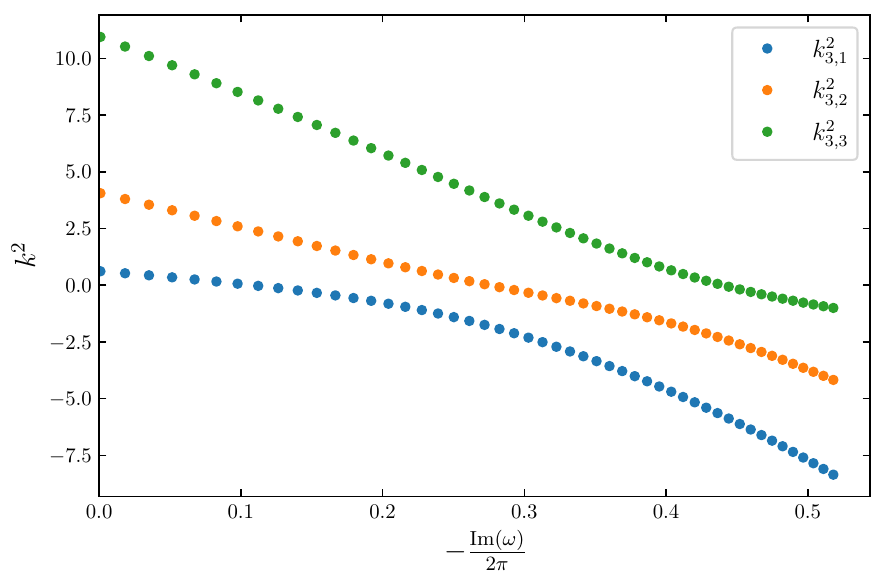}
        \caption{$n=3$}
        \label{}
    \end{subfigure}
    \begin{subfigure}{0.3\textwidth}
        \centering
        \includegraphics[width=\textwidth]{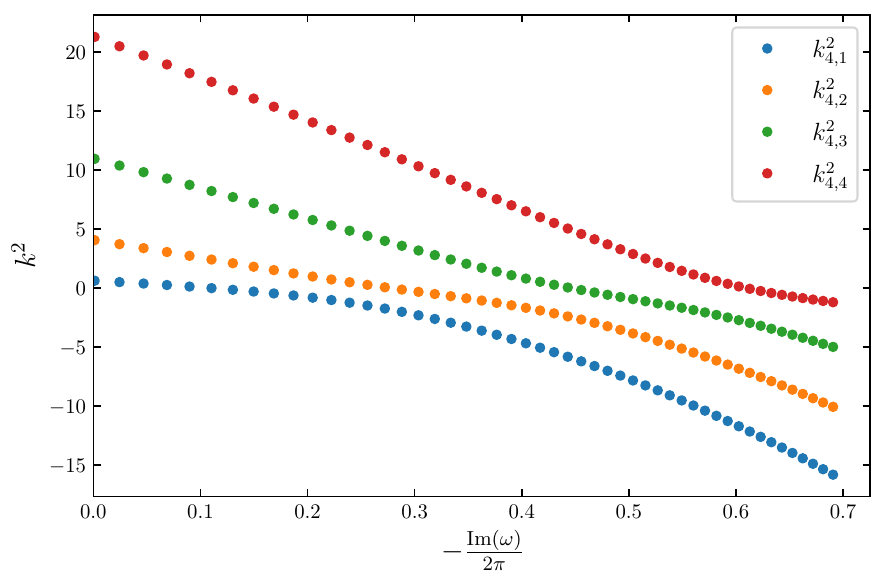}
        \caption{$n=4$}
        \label{}
    \end{subfigure}
    \begin{subfigure}{0.3\textwidth}
        \centering
        \includegraphics[width=\textwidth]{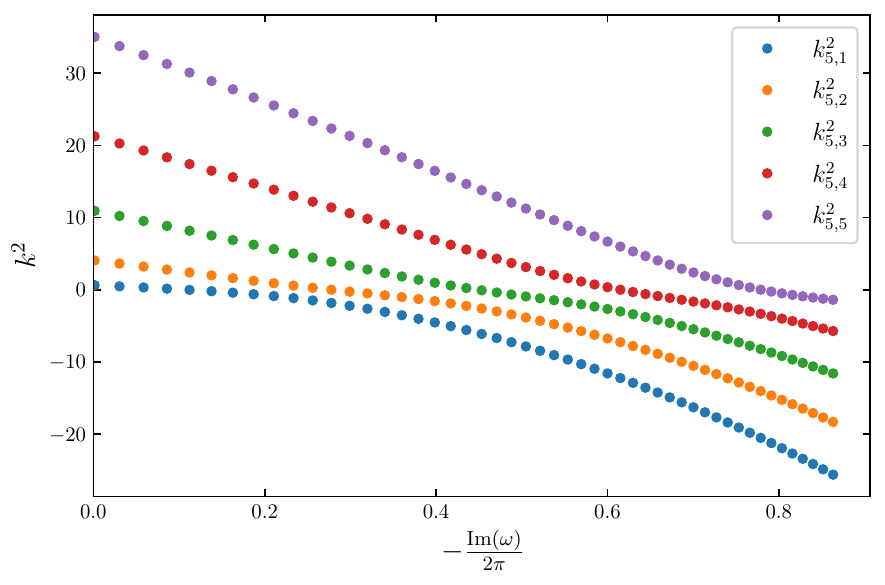}
        \caption{$n=5$}
        \label{}
    \end{subfigure}
    \begin{subfigure}{0.3\textwidth}
        \centering
        \includegraphics[width=\textwidth]{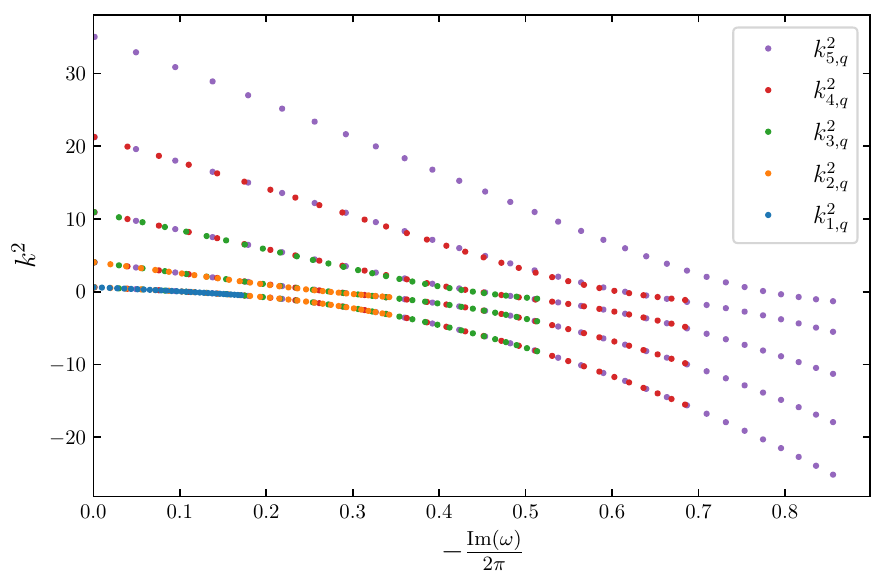}
        \caption{$n=1,2,3,4,5$}
        \label{}
    \end{subfigure}

    \caption{Finite-temperature pole-skipping trajectories $k^2(\mathrm{Im}(\omega))$ for orders $n=1$ through $n=5$. Panels (a)--(e) show individual orders; panel (f) overlays all curves, demonstrating their near-coincidence.}
    \label{fig:dispersion}
\end{figure*}

Fig.~\ref{fig:dispersion} displays the finite-temperature pole-skipping trajectories $k^2(\mathrm{Im}(\omega))$ for different orders $n$. Each curve is obtained by fixing $n$ and varying $T$ continuously; since $\omega_n = 2\pi n T$, replacing the horizontal axis from $T$ to $\mathrm{Im}(\omega_n)$ amounts to multiplying by $2\pi n$, so different orders $n$ correspond to different degrees of stretching along the horizontal axis. Each plotted point is a genuine Matsubara pole-skipping point at the corresponding temperature; no off-Matsubara continuation is performed. Notably, panel (f) shows that all curves nearly coincide after this rescaling, indicating that pole-skipping trajectories of all orders are approximately universal in the near-extremal regime. This universality is a direct consequence of the near-horizon structure: in the extremal limit, the pole-skipping momenta $k_{n,q}^2$ given by Eq.~\eqref{eq:extremal-momenta} depend on $q$ but not on $n$, so curves of different orders are related simply by the rescaling of $\omega_n = 2\pi n T$. The near-coincidence of these trajectories therefore provides direct evidence that the dynamics of high-order pole-skipping in this model is governed by near-horizon geometry rather than the detailed bulk structure at finite radial distance.

\begin{figure*}[htbp]
    \centering
    \begin{subfigure}{0.3\textwidth}
        \centering
        \includegraphics[width=\textwidth]{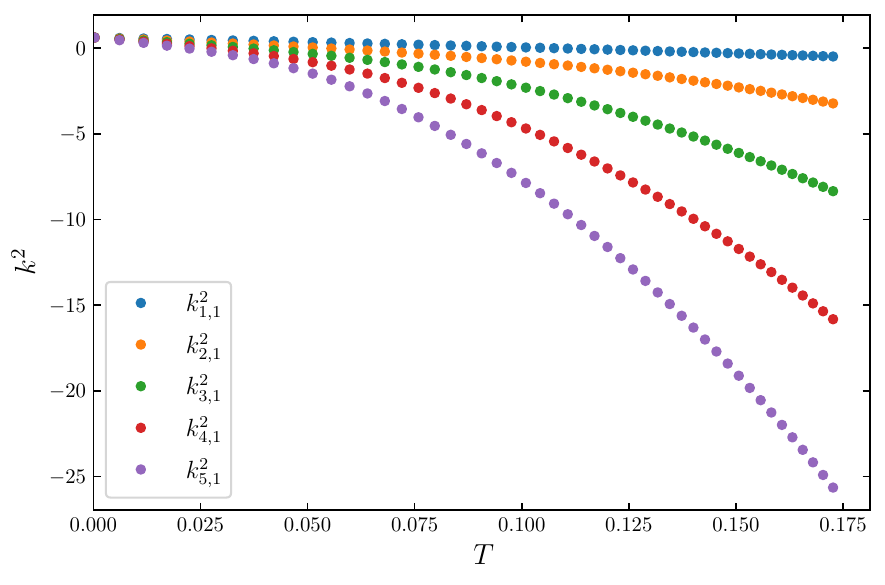}
        \caption{$q=1$}
        \label{}
    \end{subfigure}
    \begin{subfigure}{0.3\textwidth}
        \centering
        \includegraphics[width=\textwidth]{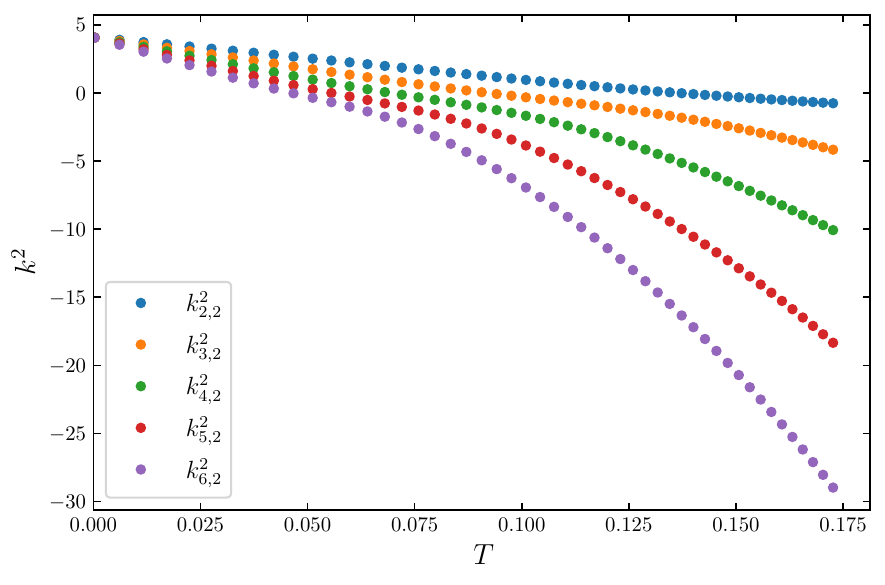}
        \caption{$q=2$}
        \label{}
    \end{subfigure}
    \begin{subfigure}{0.3\textwidth}
        \centering
        \includegraphics[width=\textwidth]{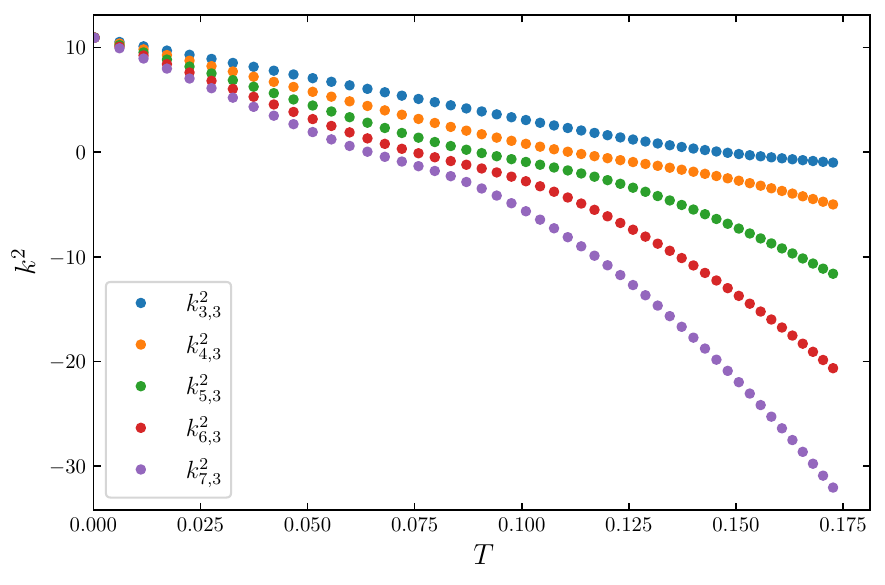}
        \caption{$q=3$}
        \label{}
    \end{subfigure}
    \begin{subfigure}{0.3\textwidth}
        \centering
        \includegraphics[width=\textwidth]{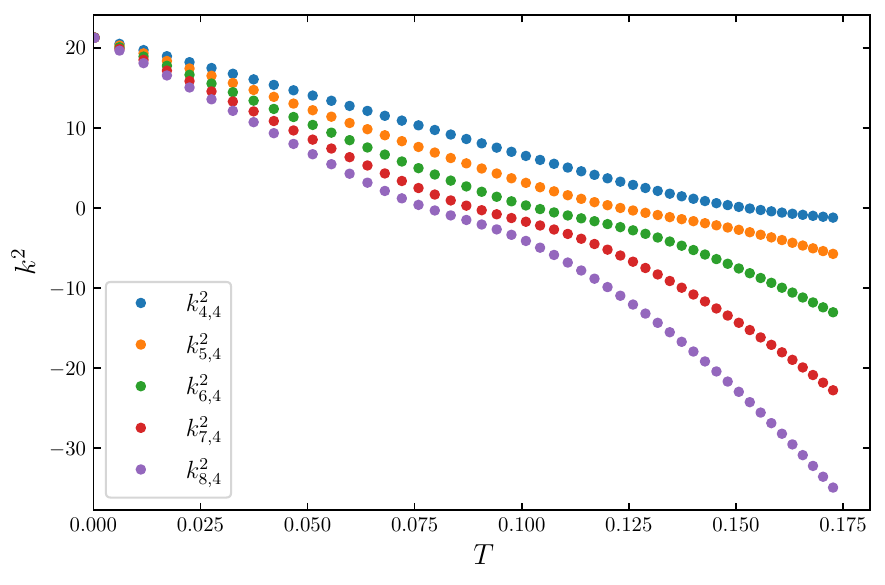}
        \caption{$q=4$}
        \label{}
    \end{subfigure}
    \begin{subfigure}{0.3\textwidth}
        \centering
        \includegraphics[width=\textwidth]{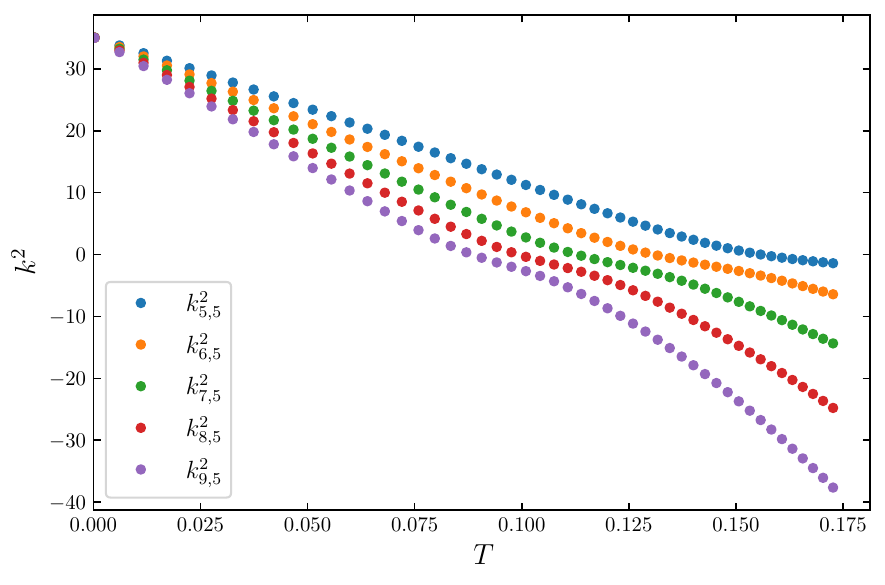}
        \caption{$q=5$}
        \label{}
    \end{subfigure}
    \begin{subfigure}{0.3\textwidth}
        \centering
        \includegraphics[width=\textwidth]{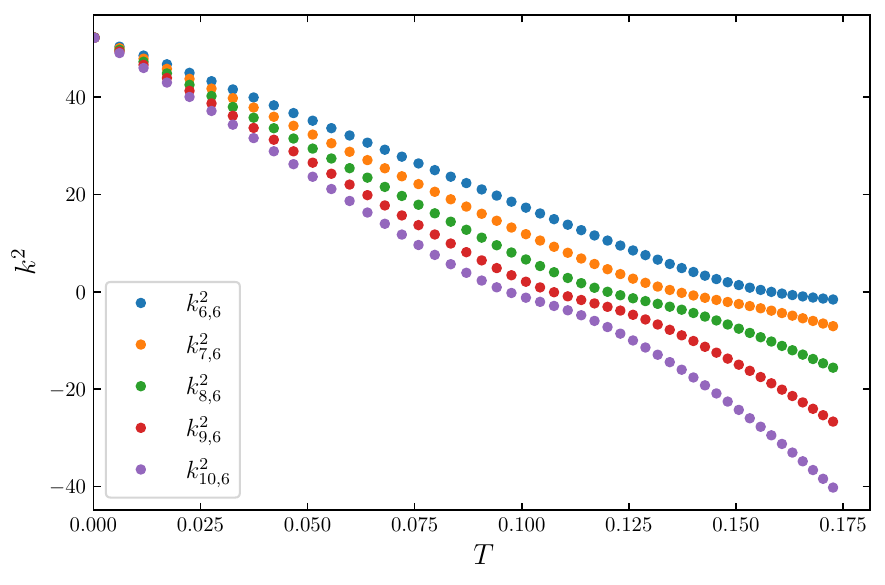}
        \caption{$q=6$}
        \label{}
    \end{subfigure}
    \caption{Temperature dependence of $k^2$ for different orders $n$ at fixed mode indices $q = 1, 2, 3, 4, 5, 6$.}
    \label{fig:k2-fixed-q}
\end{figure*}

Fig.~\ref{fig:k2-fixed-q} groups pole-skipping points by mode index $q$. At fixed $q$, higher-order points exhibit steeper temperature dependence as $n$ increases. According to Eq.~\eqref{eq:T-correction}, the slope $dk^2/dT|_{T\to 0}$ equals the correction coefficient $C_{n,q}$. Since $C_{n,q}$ increases with $n$ at fixed $q$ (as derived in Appendix~\ref{sec:app}), the increasing steepness seen in Fig.~\ref{fig:k2-fixed-q} provides a direct visual verification of the analytic structure of the temperature correction formula. Furthermore, Fig.~\ref{fig:k2-fixed-q} reveals that in the near-extremal regime ($T\to 0$), all curves exhibit an approximately linear dependence on temperature, whereas at higher temperatures a pronounced nonlinear behavior becomes evident, reflecting the growing importance of higher-order thermal corrections. A key observation from Fig.~\ref{fig:k2-fixed-q} is that, at fixed $q$, curves corresponding to different orders $n$ all converge to the same limiting value $k_q^2$ as $T \to 0$. This degeneracy, whereby pole-skipping points at distinct Matsubara orders $n$ share a common momentum in the extremal limit, follows directly from Eq.~\eqref{eq:extremal-momenta}, which depends on $q$ but not on $n$. It reflects the central result of this work: in the near-extremal limit, the pole-skipping structure is governed entirely by the near-horizon geometry, and the order $n$ enters only through the subleading temperature correction.

\begin{figure*}[htbp]
    \centering
    \begin{subfigure}{0.3\textwidth}
        \centering
        \includegraphics[width=\textwidth]{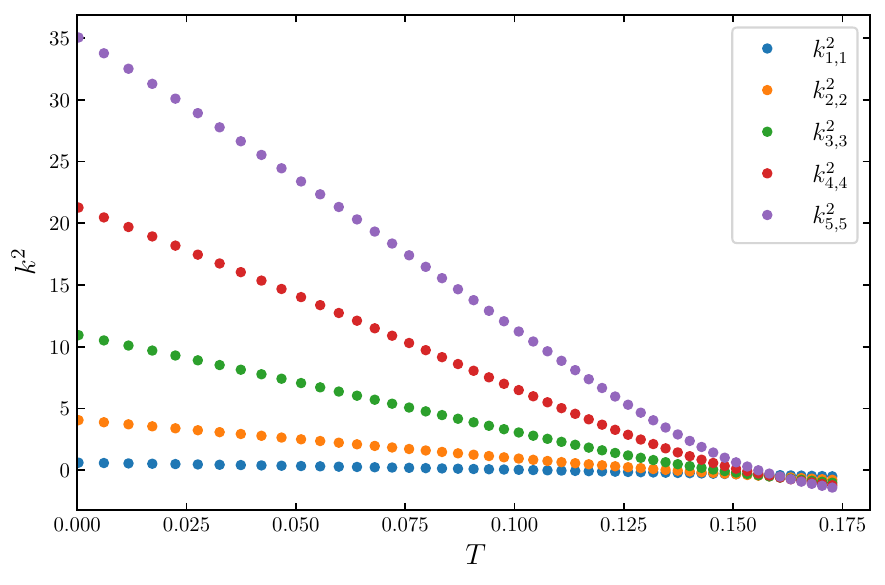}
        \caption{$j=1$}
        \label{}
    \end{subfigure}
    \begin{subfigure}{0.3\textwidth}
        \centering
        \includegraphics[width=\textwidth]{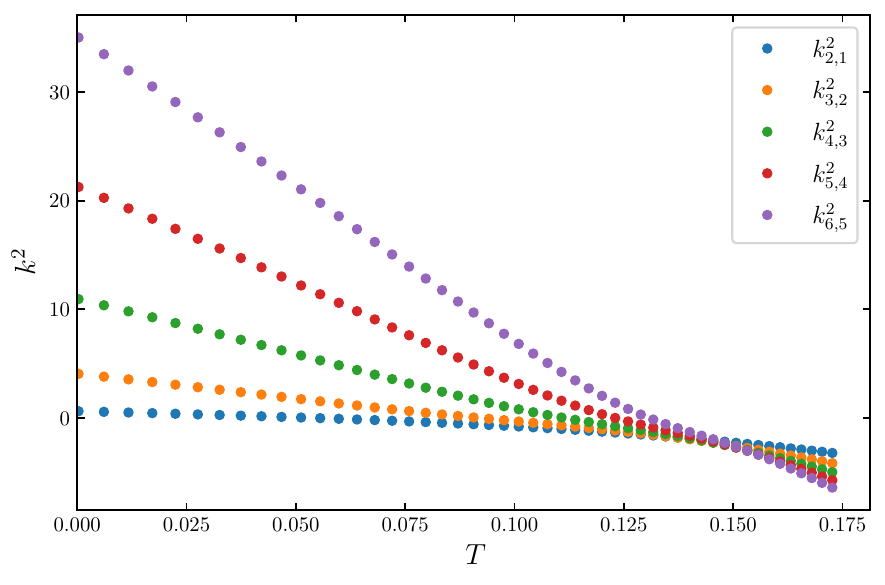}
        \caption{$j=2$}
        \label{}
    \end{subfigure}
    \begin{subfigure}{0.3\textwidth}
        \centering
        \includegraphics[width=\textwidth]{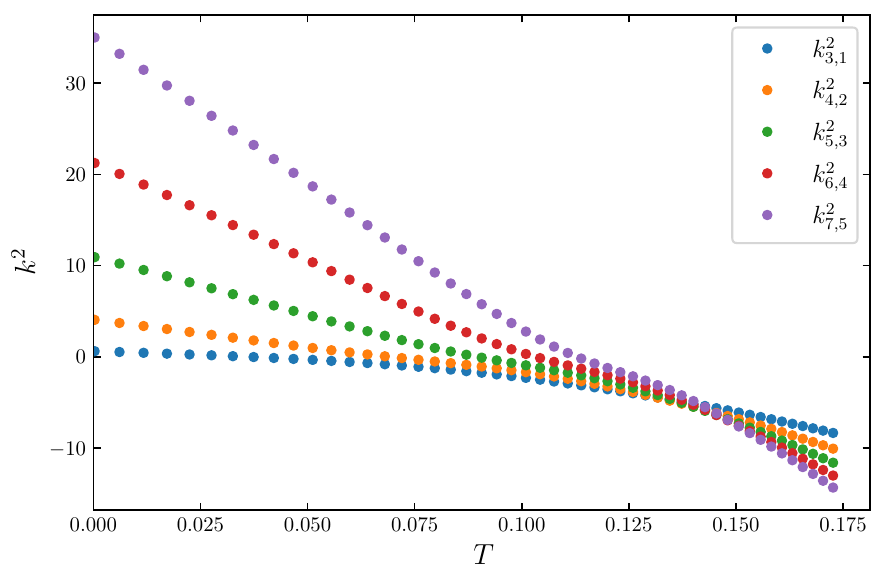}
        \caption{$j=3$}
        \label{}
    \end{subfigure}
    \begin{subfigure}{0.3\textwidth}
        \centering
        \includegraphics[width=\textwidth]{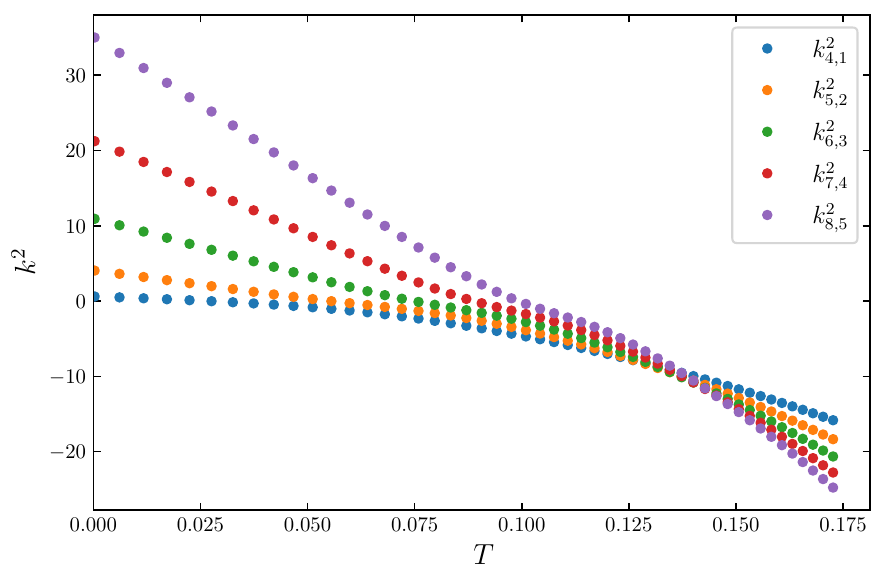}
        \caption{$j=4$}
        \label{}
    \end{subfigure}
    \begin{subfigure}{0.3\textwidth}
        \centering
        \includegraphics[width=\textwidth]{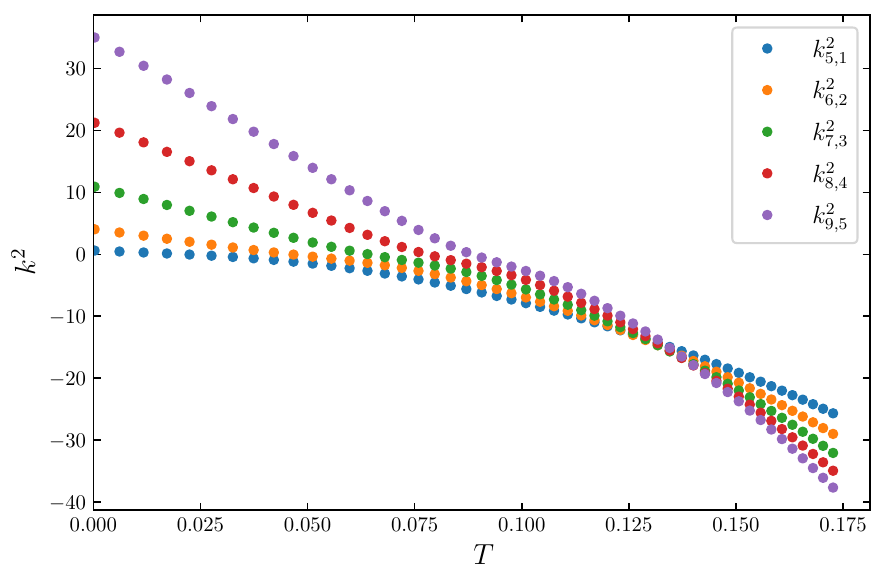}
        \caption{$j=5$}
        \label{}
    \end{subfigure}
    \begin{subfigure}{0.3\textwidth}
        \centering
        \includegraphics[width=\textwidth]{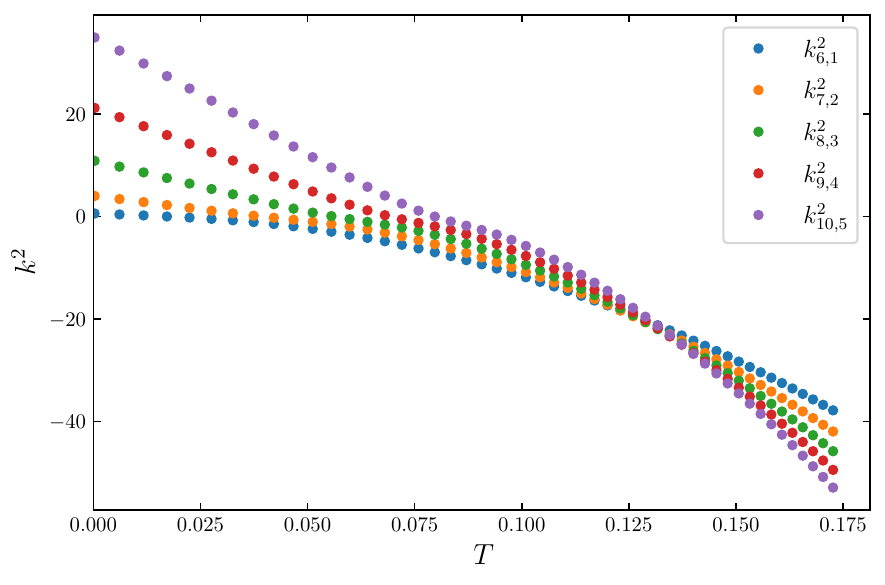}
        \caption{$j=6$}
        \label{}
    \end{subfigure}

    \caption{Temperature dependence of $k_{n+j,n}^2(T)$ for $j = 1$ through $j=6$. In each panel, four values of $n \ge j$ are shown. Curves with the same $j$ intersect at a common temperature.}
    \label{fig:k2-fixed-j}
\end{figure*}

An interesting feature emerges when examining pole-skipping points with fixed separation $j = n - q$. Fig.~\ref{fig:k2-fixed-j} shows that the functions $k_{n+j,n}^2(T)$ for different $n$ intersect at a common temperature, with the intersection point depending on $j$.

\begin{figure*}[htbp]
    \centering
    \begin{subfigure}{0.3\textwidth}
        \centering
        \includegraphics[width=\textwidth]{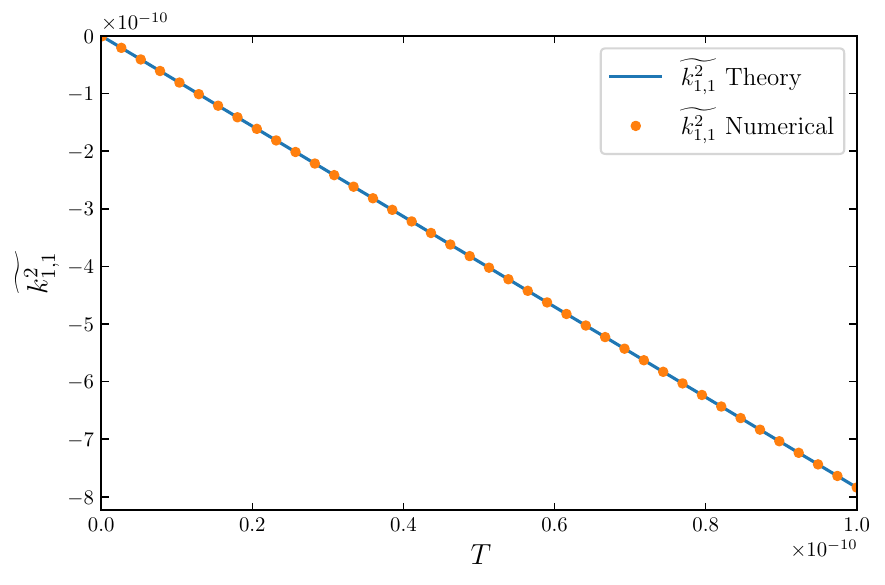}
        \caption{$(n,q)=(1,1)$}
        \label{}
    \end{subfigure}
    \begin{subfigure}{0.3\textwidth}
        \centering
        \includegraphics[width=\textwidth]{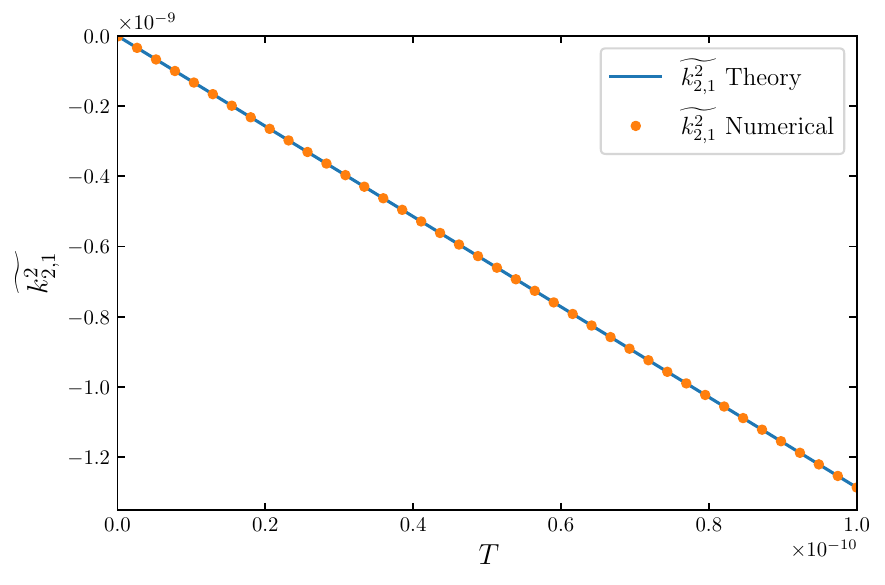}
        \caption{$(n,q)=(2,1)$}
        \label{}
    \end{subfigure}
    \begin{subfigure}{0.3\textwidth}
        \centering
        \includegraphics[width=\textwidth]{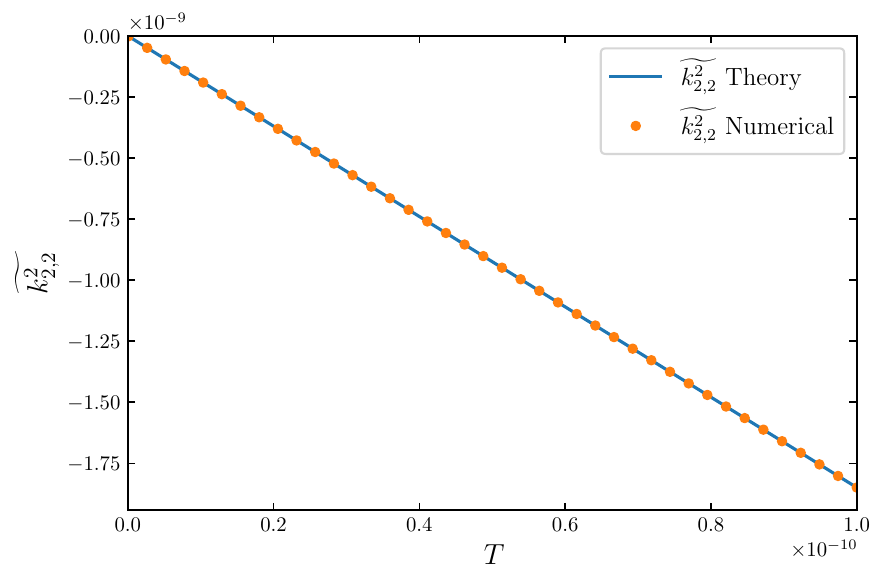}
        \caption{$(n,q)=(2,2)$}
        \label{}
    \end{subfigure}
    \begin{subfigure}{0.3\textwidth}
        \centering
        \includegraphics[width=\textwidth]{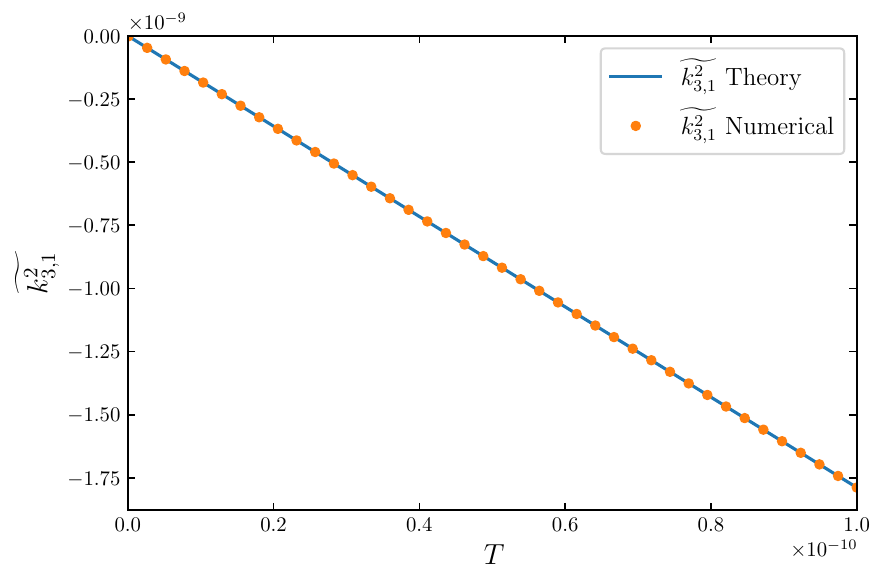}
        \caption{$(n,q)=(3,1)$}
        \label{}
    \end{subfigure}
    \begin{subfigure}{0.3\textwidth}
        \centering
        \includegraphics[width=\textwidth]{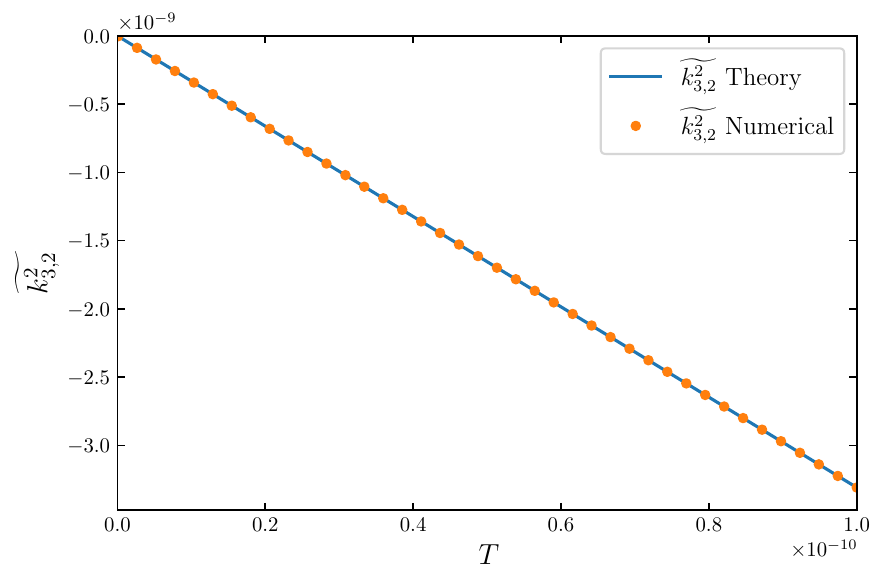}
        \caption{$(n,q)=(3,2)$}
        \label{}
    \end{subfigure}
    \begin{subfigure}{0.3\textwidth}
        \centering
        \includegraphics[width=\textwidth]{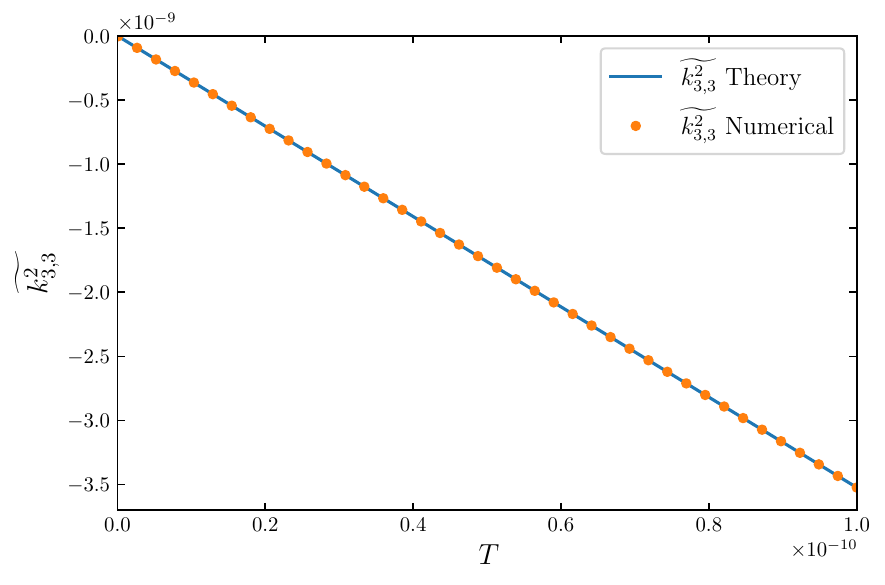}
        \caption{$(n,q)=(3,3)$}
        \label{}
    \end{subfigure}

    \caption{Comparison of analytic prediction (dashed lines) and numerical results (points) for $\widetilde{k_{n,q}^2}$ versus $T$ at various $(n,q)$ values. The agreement validates Eq.~\eqref{eq:T-correction}.}
    \label{fig:T-correction-fit}
\end{figure*}

To quantitatively test the temperature correction formula Eq.~\eqref{eq:T-correction}, we compare numerical solutions of the full determinant condition Eq.~\eqref{eq:ps-condition} with the analytic prediction. Defining $\widetilde{k_{n,q}^2}=k_{n,q}^2+m^2 h(r_h) - \frac{q(q-1)}{2}h(r_h)f''(r_h)$ to isolate the $T$-independent part, Fig.~\ref{fig:T-correction-fit} shows $\widetilde{k_{n,q}^2}$ versus $T$ for several $(n,q)$ pairs. The numerical data (points) agree well with the linear fit (dashed lines) predicted by Eq.~\eqref{eq:T-correction}, confirming that the temperature dependence is indeed linear at low $T$ for all orders and mode indices tested.

\sisetup{
table-number-alignment=center,
round-mode=places,
round-precision=6
}

\begin{table*}[t]
\centering
\begin{tabular}{c
S[table-format=2.6]
S[table-format=2.6]
S[table-format=2.6]
S[table-format=2.6]
S[table-format=2.6]
S[table-format=2.6]}
\toprule
& \multicolumn{2}{c}{Numerical} 
& \multicolumn{2}{c}{Analytic}
& \multicolumn{2}{c}{Relative\ error} \\
$(n,q)$ & {$a$} & {$-b$} & {$a$} & {$-b$} & {$a\,(10^{-29})$} & {$b\,(10^{-15})$} \\
\midrule
(1,1) & 0.619093 & 7.836716 & 0.619093 & 7.836716 & 3.516681300327846 & 4.30675047679398 \\
(2,1) & 0.619093 & 12.860460 & 0.619093 & 12.860460 & 6.809092839336182 & 4.972516330925938 \\
(2,2) & 4.068583 & 18.486404 & 4.068583 & 18.486404 & 1.104353072815411 & 3.843596309260627 \\
(3,1) & 0.619093 & 17.884204 & 0.619093 & 17.884204 & 12.29644540435015 & 6.555480680304751 \\
(3,2) & 4.068583 & 33.089127 & 4.068583 & 33.089127 & 1.8381468232060934 & 3.6505123405801005 \\
(3,3) & 10.967563 & 35.230545 & 10.967563 & 35.230545 & 0.8075991852911004 & 3.630306986234587 \\
\bottomrule
\end{tabular}
\caption{Comparison of numerical and analytic coefficients in the linear expansion 
$k_{n,q}^2 = a + bT$ near extremality, evaluated over $T\in[10^{-18},10^{-15}]$. 
Numerical calculations are performed with 100-digit precision. 
The relative error is defined as $|x_{\rm num}-x_{\rm ana}|/|x_{\rm ana}|$. 
The agreement confirms Eq.~\eqref{eq:T-correction}.}\label{tab:linear-fit}
\end{table*}

Table~\ref{tab:linear-fit} provides a quantitative comparison of the coefficients $a$ and $b$ in the linear fit $k_{n,q}^2 = a + bT$. The numerical values extracted from the full determinant calculation match the analytic prediction with relative errors at the level of $10^{-29}$--$10^{-15}$. This essentially exact agreement simultaneously validates two independent predictions: the extremal-limit momenta given by Eq.~\eqref{eq:extremal-momenta}, extracted from the intercepts $a$, and the temperature correction coefficients $C_{n,q}$ given by Eq.~\eqref{eq:T-correction}, extracted from the slopes $b$. Taken together, the results in this section confirm that the hierarchical near-extremal expansion developed in Sec.~\ref{sec:near-extremal} accurately describes high-order pole-skipping in the dyonic Gubser--Rocha model across multiple orders $n$, mode indices $q$, and temperature regimes.

\section{Conclusion}\label{sec:conclusion}

We have developed a systematic analytic framework for studying high-order pole-skipping in near-extremal holographic black holes. The central insight is that the low-temperature regime introduces a natural temperature-graded hierarchy in the near-horizon recursion relations, which drastically simplifies the determination of pole-skipping points at arbitrary order. This simplification has a concrete physical origin: as $T \to 0$, the near-horizon geometry develops an emergent AdS$_2$ throat, and the pole-skipping spectrum is reorganized by the infrared conformal structure of this AdS$_2$/CFT$_1$ geometry rather than the full bulk spacetime.

By reorganizing the standard Frobenius expansion according to powers of temperature, we have shown that the near-horizon coefficients $\Phi_j$ exhibit a power-law hierarchy $\Phi_j \sim T^{-j}\Phi_0$ . This hierarchical structure reduces the $n$-th order pole-skipping condition from an intractable $n \times n$ determinant to a factorized algebraic equation involving only near-horizon geometric data. Each pole-skipping momentum is determined by the vanishing of a single diagonal element $M_{qq}^{(0)} = 0$, which depends on the mode index $q$ but not on the order $n$. This mechanism produces a high degeneracy in the extremal limit: pole-skipping points at all orders $n \geq q$ that share the same mode index $q$ collapse onto identical momenta, a dramatic simplification from the generically $n$-dependent structure at finite temperature. The resulting formula, Eq.~\eqref{eq:extremal-momenta}, provides explicit expressions for these degenerate pole-skipping momenta and reveals several universal features: all momenta organize into discrete branches determined solely by near-horizon geometry and the scalar field mass, and include a universal branch appearing at every Matsubara frequency. We have further shown that these geometric quantities can be reexpressed in terms of black hole thermodynamic variables, including entropy density $s$, specific heat $c$, and the extremal specific heat scale $\kappa_0$, providing a thermodynamic interpretation of the pole-skipping structure and clarifying its connection to near-horizon thermal response.

The largest-momentum pole-skipping points ($q = n$) form a sequence of discrete points in the $(n, k_{n,n})$ plane. In the limit $nT \to 0$, the leading momenta grow as $k_{n,n} \propto n$, with the proportionality constant controlled entirely by near-horizon geometry. This suggests that high-order pole-skipping points encode universal features governed by horizon physics rather than bulk dynamics.

We have further computed the leading temperature corrections away from extremality and obtained analytic control over the pole-skipping trajectories in the near-extremal regime. The correction coefficient $C_{n,q}$ in Eq.~\eqref{eq:T-correction} encodes how different pole-skipping branches respond to thermal perturbations and provides a systematic expansion parameter for finite-temperature effects.

A further structural feature emerges when examining pole-skipping points with fixed separation $j = n - q$: the functions $k_{n+j,n}^2(T)$ for different values of $n$ intersect at a common temperature that depends only on $j$. This intersection pattern provides an additional organizational principle within the near-extremal pole-skipping spectrum and further supports the interpretation that the temperature structure is controlled by near-horizon data.

As a consistency check, we verified that Eqs.~\eqref{eq:extremal-momenta} and~\eqref{eq:T-correction} correctly reproduce the known pole-skipping momenta in the BTZ and planar AdS-Schwarzschild backgrounds reported in the literature. As a nontrivial verification, we applied our method to the Dyonic Gubser--Rocha model, which incorporates finite charge density and momentum relaxation. Numerical evaluation of the full determinant condition confirms the analytic predictions with excellent precision across multiple orders and branches. The agreement between numerical results and analytic formulas validates both the extremal limit expressions and the temperature correction terms, demonstrating that the hierarchical structure uncovered in this work captures the essential physics of high-order pole-skipping.

The near-coincidence of pole-skipping trajectories for different orders in the near-extremal regime provides further evidence that pole-skipping phenomena in this limit are governed by near-horizon data rather than the full bulk geometry. This universality suggests a deeper organizing principle underlying holographic Green's functions at low temperature.

Our results open several directions for future investigation. First, the method developed here is broadly applicable to general holographic black hole backgrounds, and it would be interesting to explore whether the universal features identified in this work persist across different models. Second, a main consequence of this work is that the mode index $q$ is identified with the IR conformal dimension $\Delta_{\mathrm{IR}} = q$ in the emergent near-horizon AdS$_2$ throat: the factorized pole-skipping condition selects precisely those momenta at which the scalar field matches an integer IR scaling dimension, and the discrete pole-skipping spectrum reflects the conformal tower of IR scaling dimensions in the AdS$_2$/CFT$_1$ correspondence. Whether an analogous AdS$_2$ interpretation extends to the gravitational sound channel, where the leading pole-skipping point encodes the Lyapunov exponent and butterfly velocity, remains an important open question; the structure of higher-spin perturbations in the near-AdS$_2$ throat may require a more refined analysis involving the Schwarzian mode and its couplings to matter. Third, extending the analysis to include higher-order temperature corrections would allow systematic exploration of the crossover between near-extremal and finite-temperature regimes. Finally, the connection between pole-skipping and quantum chaos suggests that the near-extremal structure uncovered here may have implications for understanding thermalization and scrambling dynamics in strongly coupled systems at low temperature.

\begin{acknowledgments}
We would like to thank Jun Nian, Keun-Young Kim, Bum-Hoon Lee, Sangjin Sin, Shaofeng Wu, Yuqi Lei for helpful discussions. This project was partially supported by the National Natural Science Foundation of China (NSFC) (Grant Nos. 12275166 and 12311540141).
\end{acknowledgments}

\appendix
\section{Iterative Near-Horizon Recursion}\label{app:iterative}

The extremal factorization condition~\eqref{eq:factorized} can also be derived via the near-horizon recursion relations. This derivation additionally yields the coefficient scaling $\Phi_j \sim T^{-j}\Phi_0$, which is used in Appendix~\ref{sec:app}.

For $1 \leq j < n$, assume
\begin{equation}
M_{jj}^{(0)} \neq 0 .
\label{assumption}
\end{equation}
For the first-order equation of motion $S_1$, we have
\begin{equation}
	S_1=\left(M_{11}^{(0)}+\mathcal{O}(T)\right)\Phi_0 - 2\pi T(n-1)\Phi_1 = 0,
\end{equation}
which gives
\begin{equation}
	\Phi_1 = \left(\frac{M_{11}^{(0)}}{2\pi(n-1)} \frac{1}{T}+ \mathcal{O}(T^0)\right)\Phi_0.
\end{equation}
Substituting this result into $S_2$
\begin{align}
	S_2&=M_{21}\Phi_0 + M_{22}\Phi_1 - 4\pi T(n-2)\Phi_2 \nonumber\\
    &=\mathcal{O}(T^0)\Phi_0+\mathcal{O}(T^0)\Phi_1+\mathcal{O}(T^1)\Phi_2 = 0.
\end{align}
Keeping only the leading terms in $T$,
\begin{equation}
	S_2=M_{22}^{(0)}\Phi_1 - 4\pi T(n-2)\Phi_2 = 0.
\end{equation}
When $h(r)$ and $f''(r)$ are regular at the horizon, we can proceed iteratively up to $S_j$, which gives
\begin{align}
	S_j&=M_{jj}^{(0)}\Phi_{j-1} - 2\pi T\, j(n-j)\Phi_j = 0,\\
    \Phi_j &\sim T^{-j} \Phi_0.
\end{align}
This yields the iterative relation
\begin{equation}\label{eq:iterative-Phi}
	\Phi_{j}=\frac{M_{jj}^{(0)}}{2\pi j(n-j)} \frac{1}{T}\Phi_{j-1}
	=\left(\prod_{\sigma=1}^j \frac{M_{\sigma\sigma}^{(0)}}{2\pi\sigma(n-\sigma)}\right)\frac{1}{T^j}\Phi_{0}.
\end{equation}

At $j=n$, the superdiagonal element vanishes, $M_{n,n+1} = 0$. The $n$-th recursion relation becomes
\begin{equation}\label{n-th order EOM}
	S_{n}=M_{nn}^{(0)}\Phi_{n-1} + M_{n,n-1}\Phi_{n-2} + \cdots = 0.
\end{equation}
Substituting Eq.~\eqref{eq:iterative-Phi} into $S_n$,
\begin{align}
    S_{n}=& M_{nn}^{(0)}\left(\prod_{\sigma=1}^{n-1} \frac{M_{\sigma\sigma}^{(0)}}{2\pi\sigma(n-\sigma)}\right)\frac{1}{T^{n-1}}\Phi_{n-1} \nonumber \\
    &+ \mathcal{O}(T^{-(n-2)})\Phi_{0} + \cdots = 0.
\end{align}
For a non-trivial solution with two independent parameters $(\Phi_0, \Phi_n)$, the coefficient of $\Phi_0$ must vanish, reproducing the extremal factorization condition~\eqref{eq:factorized}.

When $M_{q,q}^{(0)} = 0$, the iterative relation~\eqref{eq:iterative-Phi} remains valid for $j < q$ but breaks down at $j \geq q$. In this case the recursion reduces to
\begin{equation}\label{new iterative relation}
	\Phi_{j}
	=\left(\prod_{\sigma=1}^j \frac{M_{\sigma\sigma}^{(0)}}{2\pi\sigma(n-\sigma)}\right)\frac{1}{T^{j-1}}\Phi_{0}.
\end{equation}

At $j=n$, the superdiagonal element vanishes $M_{n,n+1} = 0$.
The $n$-th recursion relation becomes
\begin{equation}\label{n-th order EOM}
	S_{n}=M_{nn}^{(0)}\Phi_{n-1} + M_{n,n-1}\Phi_{n-2} + \cdots = 0.
\end{equation}
Substituting Eq.~\eqref{eq:iterative-Phi} into $S_n$
\begin{align}
    S_{n}=& M_{nn}^{(0)}\left(\prod_{\sigma=1}^{n-1} \frac{M_{\sigma\sigma}^{(0)}}{2\pi\sigma(n-\sigma)}\right)\frac{1}{T^{n-1}}\Phi_{n-1} \nonumber \\
    &+ \mathcal{O}(T^{-(n-2)})\Phi_{0} + \cdots = 0
\end{align}

For a non-trivial solution to exist with two independent parameters $(\Phi_0, \Phi_n)$, we require the coefficient of $\Phi_0$ to vanish; this reduces to the extremal factorization condition:
\begin{equation}\label{eq:factorized}
	\prod_{\sigma=1}^n M_{\sigma\sigma}^{(0)} = 0.
\end{equation}

When $M_{q,q}^{(0)} = 0$, the iterative relation Eq.~\eqref{eq:iterative-Phi} remains valid for $j < q$, but breaks down at $j \geq q$ where the diagonal element vanishes. In this case, the recursion relation reduces to
\begin{equation}\label{new iterative relation}
	\Phi_{j}
	=\left(\prod_{\sigma=1}^j \frac{M_{\sigma\sigma}^{(0)}}{2\pi\sigma(n-\sigma)}\right)\frac{1}{T^{j-1}}\Phi_{0}.
\end{equation}

This also implies that the matrix in Eq.~\eqref{condition} can be reduced to an upper bidiagonal form
\begin{widetext}
    \begin{equation}
               \begin{aligned}
       &\mathcal{M}(\omega,k^2)\cdot \Phi\equiv
       \begin{pmatrix}
       M_{11}^{(0)} & (2\pi T+i\omega) & 0    & 0  &\cdots\\
       0 & M_{22}^{(0)}& (4\pi T+i\omega)& 0   &\cdots\\
       0 & 0&  M_{33}^{(0)}&(6\pi T+i\omega) &\cdots\\
           \vdots   &  \vdots&  \vdots  &\vdots   &\ddots\\
       \end{pmatrix}
       \times\begin{pmatrix}
           \Phi_0\\
           \Phi_1\\
           \Phi_2 \\
           \vdots \\
       \end{pmatrix}=0\,.
           \end{aligned}
           \label{condition_bidiag}
    \end{equation}
    \end{widetext}
    
\section{Linear temperature corrections in Pole-Skipping}\label{sec:app}
We find that $M_{ij}$ can be decomposed into
\begin{equation}
   M_{ij} = M_{ij}^0 + M_{ij}^1 T .
\end{equation}
Only $M_{j,j}$ depends explicitly on $k$.

We consider $k_{n,q}^2$ and the $q$th-order equation of motion
   \begin{align}
   S_q &= 0= M_{q,q+1} \Phi_q+M_{q,q}\Phi_{q-1}+M_{q,q-1}\Phi_{q-2} +\cdots \nonumber \\
&\approx 	M_{q,q+1}^1 T \Phi_q+\left(-\mathit{C_{n,q}}+M_{q,q}^1\right)T\Phi_{q-1}+M_{q,q-1}^0\Phi_{q-2}
\end{align}
where $0<q< n$. For the first $(q-1)$th-order equations of motion, Eq.~\eqref{eq:iterative-Phi} remains valid.
Substituting Eq.~\eqref{eq:iterative-Phi}, we obtain
\begin{equation}
   \Phi_{q} = -\frac{1}{M_{q,q+1}^1}\left(-\mathit{C_{n,q}}+M_{q,q}^1-\frac{M_{q,q-1}^0 
    M_{q-1,q}^1}{M_{q-1,q-1}^0}\right)\Phi_{q-1} .
\end{equation}

Here, $\Phi_{q}$ and $\Phi_{q-1}$ are of the same order.
We further consider $(q+1)$th-order equation of motion
\begin{equation}
   S_{q+1} = 0= M_{q+1,q+2} \Phi_{q+1}+M_{q+1,q+1}\Phi_{q}+M_{q+1,q}\Phi_{q-1}+\cdots
\end{equation}
The extremal factorization condition, Eq.~\eqref{eq:factorized}, requires that
\begin{equation}
   -\frac{M_{q+1,q+1}^0}{M_{q,q+1}^1}\left(-\mathit{C_{n,q}}+M_{q,q}^1-\frac{M_{q,q-1}^0 
    M_{q-1,q}^1}{M_{q-1,q-1}^0}\right)+M_{q+1,q}^0 = 0 .
\end{equation}
Thus, we obtain
\begin{equation}
   \mathit{C_{n,q}} = M_{q,q}^1 - \frac{M_{q+1,q+1}^0 M_{q+1,q}^0}{M_{q,q+1}^1}
    - \frac{M_{q,q-1}^0  M_{q-1,q}^1}{M_{q-1,q-1}^0} .
\end{equation}	
Substituting the explicit values, we obtain
\begin{widetext}
\begin{equation}
\begin{split}
   \mathit{C_{n,q}} = & \frac{2 \pi}{3 f''(r)} \Bigg[ 3 h'(r) \Big( (q-1) (q (d+2 q-2)-n (d+2 q-1)) f''(r)  + 2 m^2 (2 n-2 q+1) \Big) \\
   & + f^{(3)}(r) h(r) \Big( -2 n (q-1) q+n + (2 q-3) q^2+q \Big) \Bigg]
\end{split}
\end{equation}
\end{widetext}
Moreover, when $q=1$,
\begin{equation}
   \mathit{C_{n,1}} = M_{1,1}^1 - \frac{M_{2,2}^0 M_{2,1}^0}{M_{1,2}^1}.
\end{equation}
When $q=n$,
\begin{equation}
   \mathit{C_{n,n}} = M_{n,n}^1 - \frac{M_{n,n-1}^0  M_{n-1,n}^1}{M_{n-1,n-1}^0}.
\end{equation}

\bibliographystyle{apsrev4-2}

\bibliography{refs}

@article{Maldacena:1997re,
    author = "Maldacena, Juan Martin",
    title = "{The Large $N$ limit of superconformal field theories and supergravity}",
    eprint = "hep-th/9711200",
    archivePrefix = "arXiv",
    reportNumber = "HUTP-97-A097, HUTP-98-A097",
    doi = "10.4310/ATMP.1998.v2.n2.a1",
    journal = "Adv. Theor. Math. Phys.",
    volume = "2",
    pages = "231--252",
    year = "1998"
}

@article{Gubser:2008yx,
    author = "Gubser, Steven S. and Rocha, Fabio D.",
    title = "{The gravity dual to a quantum critical point with spontaneous symmetry breaking}",
    eprint = "0807.1737",
    archivePrefix = "arXiv",
    primaryClass = "hep-th",
    reportNumber = "PUPT-2289",
    doi = "10.1103/PhysRevLett.102.061601",
    journal = "Phys. Rev. Lett.",
    volume = "102",
    pages = "061601",
    year = "2009"
}

@article{Donos:2014cya,
    author = "Donos, Aristomenis and Gauntlett, Jerome P.",
    title = "{Thermoelectric DC conductivities from black hole horizons}",
    eprint = "1406.4742",
    archivePrefix = "arXiv",
    primaryClass = "hep-th",
    reportNumber = "Imperial-TP-AT-2014-03",
    doi = "10.1007/JHEP11(2014)081",
    journal = "JHEP",
    volume = "11",
    pages = "081",
    year = "2014"
}

@article{Duff:1994an,
    author = "Duff, M. J. and Khuri, Ramzi R. and Lu, J. X.",
    title = "{String solitons}",
    eprint = "hep-th/9412184",
    archivePrefix = "arXiv",
    reportNumber = "NI-94017, CTP-TAMU-67-92, MCGILL-94-53, CERN-TH-7542-94",
    doi = "10.1016/0370-1573(95)00002-X",
    journal = "Phys. Rept.",
    volume = "259",
    pages = "213--326",
    year = "1995"
}

@article{Gubser:1998bc,
    author = "Gubser, S. S. and Klebanov, Igor R. and Polyakov, Alexander M.",
    title = "{Gauge theory correlators from noncritical string theory}",
    eprint = "hep-th/9802109",
    archivePrefix = "arXiv",
    reportNumber = "PUPT-1767",
    doi = "10.1016/S0370-2693(98)00377-3",
    journal = "Phys. Lett. B",
    volume = "428",
    pages = "105--114",
    year = "1998"
}

@article{Witten:1998qj,
    author = "Witten, Edward",
    title = "{Anti de Sitter space and holography}",
    eprint = "hep-th/9802150",
    archivePrefix = "arXiv",
    reportNumber = "IASSNS-HEP-98-15",
    doi = "10.4310/ATMP.1998.v2.n2.a2",
    journal = "Adv. Theor. Math. Phys.",
    volume = "2",
    pages = "253--291",
    year = "1998"
}

@article{Grozdanov:2017ajz,
    author = "Grozdanov, Sa{\v{s}}o and Schalm, Koenraad and Scopelliti, Vincenzo",
    title = "{Black hole scrambling from hydrodynamics}",
    eprint = "1710.00921",
    archivePrefix = "arXiv",
    primaryClass = "hep-th",
    reportNumber = "MIT-CTP-4940, MIT-CTP/4940",
    doi = "10.1103/PhysRevLett.120.231601",
    journal = "Phys. Rev. Lett.",
    volume = "120",
    number = "23",
    pages = "231601",
    year = "2018"
}

@article{Blake:2016wvh,
    author = "Blake, Mike",
    title = "{Universal Charge Diffusion and the Butterfly Effect in Holographic Theories}",
    eprint = "1603.08510",
    archivePrefix = "arXiv",
    primaryClass = "hep-th",
    reportNumber = "DAMTP-2016-27",
    doi = "10.1103/PhysRevLett.117.091601",
    journal = "Phys. Rev. Lett.",
    volume = "117",
    number = "9",
    pages = "091601",
    year = "2016"
}

@article{Blake:2018leo,
    author = "Blake, Mike and Davison, Richard A. and Grozdanov, Sa{\v{s}}o and Liu, Hong",
    title = "{Many-body chaos and energy dynamics in holography}",
    eprint = "1809.01169",
    archivePrefix = "arXiv",
    primaryClass = "hep-th",
    reportNumber = "MIT-CTP/5046",
    doi = "10.1007/JHEP10(2018)035",
    journal = "JHEP",
    volume = "10",
    pages = "035",
    year = "2018"
}

@article{Blake:2017ris,
    author = "Blake, Mike and Lee, Hyunseok and Liu, Hong",
    title = "{A quantum hydrodynamical description for scrambling and many-body chaos}",
    eprint = "1801.00010",
    archivePrefix = "arXiv",
    primaryClass = "hep-th",
    reportNumber = "MIT-CTP/4975, MIT-CTP-4975",
    doi = "10.1007/JHEP10(2018)127",
    journal = "JHEP",
    volume = "10",
    pages = "127",
    year = "2018"
}

@article{Natsuume:2019sfp,
    author = "Natsuume, Makoto and Okamura, Takashi",
    title = "{Holographic chaos, pole-skipping, and regularity}",
    eprint = "1905.12014",
    archivePrefix = "arXiv",
    primaryClass = "hep-th",
    reportNumber = "KEK-TH-2128",
    doi = "10.1093/ptep/ptz155",
    journal = "PTEP",
    volume = "2020",
    number = "1",
    pages = "013B07",
    year = "2020"
}

@article{Ahn:2020bks,
    author = "Ahn, Yongjun and Jahnke, Viktor and Jeong, Hyun-Sik and Kim, Keun-Young and Lee, Kyung-Sun and Nishida, Mitsuhiro",
    title = "{Pole-skipping of scalar and vector fields in hyperbolic space: conformal blocks and holography}",
    eprint = "2006.00974",
    archivePrefix = "arXiv",
    primaryClass = "hep-th",
    doi = "10.1007/JHEP09(2020)111",
    journal = "JHEP",
    volume = "09",
    pages = "111",
    year = "2020"
}

@article{Yuan:2020fvv,
    author = "Yuan, Haiming and Ge, Xian-Hui",
    title = "{Pole-skipping and hydrodynamic analysis in Lifshitz, AdS$_{2}$ and Rindler geometries}",
    eprint = "2012.15396",
    archivePrefix = "arXiv",
    primaryClass = "hep-th",
    doi = "10.1007/JHEP06(2021)165",
    journal = "JHEP",
    volume = "06",
    pages = "165",
    year = "2021"
}

@article{Ceplak:2019ymw,
    author = "Ceplak, Nejc and Ramdial, Kushala and Vegh, David",
    title = "{Fermionic pole-skipping in holography}",
    eprint = "1910.02975",
    archivePrefix = "arXiv",
    primaryClass = "hep-th",
    reportNumber = "QMUL-PH-19-22",
    doi = "10.1007/JHEP07(2020)203",
    journal = "JHEP",
    volume = "07",
    pages = "203",
    year = "2020"
}

@article{Jansen:2020hfd,
    author = "Jansen, Aron and Pantelidou, Christiana",
    title = "{Quasinormal modes in charged fluids at complex momentum}",
    eprint = "2007.14418",
    archivePrefix = "arXiv",
    primaryClass = "hep-th",
    doi = "10.1007/JHEP10(2020)121",
    journal = "JHEP",
    volume = "10",
    pages = "121",
    year = "2020"
}

@article{Yuan:2024utc,
    author = "Yuan, Haiming and Ge, Xian-Hui and Kim, Keun-Young",
    title = "{Pole skipping in two-dimensional de Sitter spacetime and double-scaled SYK model}",
    eprint = "2408.12330",
    archivePrefix = "arXiv",
    primaryClass = "hep-th",
    doi = "10.1103/f3cb-kmnc",
    journal = "Phys. Rev. D",
    volume = "112",
    number = "2",
    pages = "026022",
    year = "2025"
}

@article{Yuan:2025ivz,
    author = "Yuan, Haiming and Ge, Xian-Hui and Kim, Keun-Young",
    title = "{Pole-skipping in the de Sitter horizon structure}",
    eprint = "2512.19087",
    archivePrefix = "arXiv",
    primaryClass = "hep-th",
    journal = {},
    month = "12",
    year = "2025"
}

@article{Yuan:2021ets,
    author = "Yuan, Haiming and Ge, Xian-Hui",
    title = "{Analogue of the pole-skipping phenomenon in acoustic black holes}",
    eprint = "2110.08074",
    archivePrefix = "arXiv",
    primaryClass = "hep-th",
    doi = "10.1140/epjc/s10052-022-10129-y",
    journal = "Eur. Phys. J. C",
    volume = "82",
    number = "2",
    pages = "167",
    year = "2022"
}

@article{Liu:2018kfw,
    author = "Liu, Hong and Glorioso, Paolo",
    title = "{Lectures on non-equilibrium effective field theories and fluctuating hydrodynamics}",
    eprint = "1805.09331",
    archivePrefix = "arXiv",
    primaryClass = "hep-th",
    reportNumber = "MIT-CTP/5018; EFI-18-8, MIT-CTP-5018, EFI-18-8",
    doi = "10.22323/1.305.0008",
    journal = "PoS",
    volume = "TASI2017",
    pages = "008",
    year = "2018"
}

@article{Yuan:2023tft,
    author = "Yuan, Haiming and Ge, Xian-Hui and Kim, Keun-Young and Ji, Chang-Woo and Ahn, Yongjun",
    title = "{Pole-skipping points in 2D gravity and SYK model}",
    eprint = "2303.04801",
    archivePrefix = "arXiv",
    primaryClass = "hep-th",
    doi = "10.1007/JHEP08(2023)157",
    journal = "JHEP",
    volume = "08",
    pages = "157",
    year = "2023"
}

@article{Blake:2019otz,
    author = "Blake, Mike and Davison, Richard A. and Vegh, David",
    title = "{Horizon constraints on holographic Green{\textquoteright}s functions}",
    eprint = "1904.12883",
    archivePrefix = "arXiv",
    primaryClass = "hep-th",
    reportNumber = "MIT-CTP/5116",
    doi = "10.1007/JHEP01(2020)077",
    journal = "JHEP",
    volume = "01",
    pages = "077",
    year = "2020"
}

@article{Ge:2023yom,
    author = "Ge, Xian-Hui and Xu, Zhaojie",
    title = "{Thermo-electric transport of dyonic Gubser-Rocha black holes}",
    eprint = "2310.12067",
    archivePrefix = "arXiv",
    primaryClass = "hep-th",
    doi = "10.1007/JHEP03(2024)069",
    journal = "JHEP",
    volume = "03",
    pages = "069",
    year = "2024"
}

@article{Xu:2023qlu,
    author = "Xu, Zhaojie",
    title = "{Duality and Triality Families of Analytic Black Hole Solutions}",
    eprint = "2312.04177",
    archivePrefix = "arXiv",
    primaryClass = "hep-th",
    journal = {},
    month = "12",
    year = "2023"
}

@article{Ishigaki:2024djz,
    author = "Ishigaki, Shuta and Xu, Zhaojie",
    title = "{Thermodynamics, magnetic properties, and global U(1) symmetry breaking of the S-type Gubser-Rocha model}",
    eprint = "2406.00666",
    archivePrefix = "arXiv",
    primaryClass = "hep-th",
    doi = "10.1007/JHEP12(2024)182",
    journal = "JHEP",
    volume = "12",
    pages = "182",
    year = "2024"
}

@article{Jeong:2023rck,
    author = "Jeong, Hyun-Sik and Ji, Chang-Woo and Kim, Keun-Young",
    title = "{Pole-skipping in rotating BTZ black holes}",
    eprint = "2306.14805",
    archivePrefix = "arXiv",
    primaryClass = "hep-th",
    reportNumber = "IFT-UAM/CSIC-23-56",
    doi = "10.1007/JHEP08(2023)139",
    journal = "JHEP",
    volume = "08",
    pages = "139",
    year = "2023"
}

@article{Pan:2024azf,
    author = "Pan, Wen-Bin and Sun, Ya-Wen and Wang, Yuan-Tai",
    title = "{Pole-skipping for massive fields and the Stueckelberg formalism}",
    eprint = "2404.17354",
    archivePrefix = "arXiv",
    primaryClass = "hep-th",
    doi = "10.1007/JHEP07(2024)256",
    journal = "JHEP",
    volume = "07",
    pages = "256",
    year = "2024"
}

@article{Lu:2025pal,
    author = "Lu, Zhenkang and Ran, Cheng and Wu, Shao-feng",
    title = "{Algebraic structure underlying pole-skipping points}",
    eprint = "2507.13306",
    archivePrefix = "arXiv",
    primaryClass = "hep-th",
    doi = "10.1103/lgkz-gyl1",
    journal = "Phys. Rev. D",
    volume = "113",
    number = "4",
    pages = "046008",
    year = "2026"
}

@article{Lu:2025jgk,
    author = "Lu, Zhenkang and Ran, Cheng and Wu, Shao-feng",
    title = "{Bulk Spacetime Encoding via Boundary Ambiguities}",
    eprint = "2506.12890",
    archivePrefix = "arXiv",
    primaryClass = "hep-th",
    doi = "10.1103/gmd3-zt39",
    journal = "Phys. Rev. Lett.",
    volume = "136",
    number = "6",
    pages = "061603",
    year = "2026"
}

@article{Abbasi:2020ykq,
    author = "Abbasi, Navid and Tahery, Sara",
    title = "{Complexified quasinormal modes and the pole-skipping in a holographic system at finite chemical potential}",
    eprint = "2007.10024",
    archivePrefix = "arXiv",
    primaryClass = "hep-th",
    doi = "10.1007/JHEP10(2020)076",
    journal = "JHEP",
    volume = "10",
    pages = "076",
    year = "2020"
}

@article{Ramirez:2020qer,
    author = "Ramirez, David M.",
    title = "{Chaos and pole skipping in CFT$_{2}$}",
    eprint = "2009.00500",
    archivePrefix = "arXiv",
    primaryClass = "hep-th",
    doi = "10.1007/JHEP12(2021)006",
    journal = "JHEP",
    volume = "12",
    pages = "006",
    year = "2021"
}

@article{Choi:2020tdj,
    author = "Choi, Changha and Mezei, M{\'a}rk and S{\'a}rosi, G{\'a}bor",
    title = "{Pole skipping away from maximal chaos}",
    eprint = "2010.08558",
    archivePrefix = "arXiv",
    primaryClass = "hep-th",
    reportNumber = "CERN-TH-2020-171",
    doi = "10.1007/JHEP02(2021)207",
    journal = "JHEP",
    volume = "02",
    pages = "207",
    year = "2021"
}

@article{Ahn:2020baf,
    author = "Ahn, Yongjun and Jahnke, Viktor and Jeong, Hyun-Sik and Kim, Keun-Young and Lee, Kyung-Sun and Nishida, Mitsuhiro",
    title = "{Classifying pole-skipping points}",
    eprint = "2010.16166",
    archivePrefix = "arXiv",
    primaryClass = "hep-th",
    doi = "10.1007/JHEP03(2021)175",
    journal = "JHEP",
    volume = "03",
    pages = "175",
    year = "2021"
}

@article{Natsuume:2021fhn,
    author = "Natsuume, Makoto and Okamura, Takashi",
    title = "{Nonuniqueness of scattering amplitudes at special points}",
    eprint = "2108.07832",
    archivePrefix = "arXiv",
    primaryClass = "quant-ph",
    reportNumber = "KEK-TH-2342",
    doi = "10.1103/PhysRevD.104.126007",
    journal = "Phys. Rev. D",
    volume = "104",
    number = "12",
    pages = "126007",
    year = "2021"
}

@article{Blake:2021hjj,
    author = "Blake, Mike and Davison, Richard A.",
    title = "{Chaos and pole-skipping in rotating black holes}",
    eprint = "2111.11093",
    archivePrefix = "arXiv",
    primaryClass = "hep-th",
    doi = "10.1007/JHEP01(2022)013",
    journal = "JHEP",
    volume = "01",
    pages = "013",
    year = "2022"
}

@article{Amrahi:2023xso,
    author = "Amrahi, B. and Asadi, M. and Taghinavaz, F.",
    title = "{Chaos near to the critical point: butterfly effect and pole-skipping}",
    eprint = "2305.00298",
    archivePrefix = "arXiv",
    primaryClass = "hep-th",
    doi = "10.1140/epjc/s10052-024-12854-y",
    journal = "Eur. Phys. J. C",
    volume = "84",
    number = "5",
    pages = "505",
    year = "2024"
}

@article{Chua:2025vig,
    author = "Chua, Wan Zhen and Hartman, Thomas and Weng, Wayne W.",
    title = "{Replica manifolds, pole skipping, and the butterfly effect}",
    eprint = "2504.08139",
    archivePrefix = "arXiv",
    primaryClass = "hep-th",
    journal = {},
    month = "4",
    year = "2025"
}

@article{Basu:2025exh,
    author = "Basu, Debarshi and Chandra, Ashish and Wen, Qiang",
    title = "{Butterfly effect and TT{\textasciimacron}-deformation}",
    eprint = "2505.14331",
    archivePrefix = "arXiv",
    primaryClass = "hep-th",
    doi = "10.1103/qq7z-vkxv",
    journal = "Phys. Rev. D",
    volume = "112",
    number = "10",
    pages = "106007",
    year = "2025"
}

@article{Gao:2025ohk,
    author = "Gao, Ping and Liu, Hong",
    title = "{Probing Stringy Horizons with Pole-Skipping in Non-Maximal Chaotic Systems}",
    eprint = "2512.20700",
    archivePrefix = "arXiv",
    primaryClass = "hep-th",
    reportNumber = "MIT-CTP/5992",
    journal = {},
    month = "12",
    year = "2025"
}

@article{Wald:1999vt,
    author = "Wald, Robert M.",
    title = "{The thermodynamics of black holes}",
    eprint = "gr-qc/9912119",
    archivePrefix = "arXiv",
    doi = "10.12942/lrr-2001-6",
    journal = "Living Rev. Rel.",
    volume = "4",
    pages = "6",
    year = "2001"
}

@article{Natsuume:2020snz,
    author = "Natsuume, Makoto and Okamura, Takashi",
    title = "{Pole-skipping and zero temperature}",
    eprint = "2011.10093",
    archivePrefix = "arXiv",
    primaryClass = "hep-th",
    reportNumber = "KEK-TH-2277",
    doi = "10.1103/PhysRevD.103.066017",
    journal = "Phys. Rev. D",
    volume = "103",
    number = "6",
    pages = "066017",
    year = "2021"
}

@article{Sachdev:2010um,
    author = "Sachdev, Subir",
    title = "{Holographic metals and the fractionalized Fermi liquid}",
    eprint = "1006.3794",
    archivePrefix = "arXiv",
    primaryClass = "hep-th",
    doi = "10.1103/PhysRevLett.105.151602",
    journal = "Phys. Rev. Lett.",
    volume = "105",
    pages = "151602",
    year = "2010"
}

@article{Faulkner:2009wj,
    author = "Faulkner, Thomas and Liu, Hong and McGreevy, John and Vegh, David",
    title = "{Emergent quantum criticality, Fermi surfaces, and AdS(2)}",
    eprint = "0907.2694",
    archivePrefix = "arXiv",
    primaryClass = "hep-th",
    reportNumber = "MIT-CTP-4050",
    doi = "10.1103/PhysRevD.83.125002",
    journal = "Phys. Rev. D",
    volume = "83",
    pages = "125002",
    year = "2011"
}

\end{document}